\documentclass[prd,twocolumn]{revtex4}
\usepackage{epsfig}
\usepackage{amsmath}
\newcommand{\be}{\begin{equation}}
\newcommand{\ee}{\end{equation}}
\newcommand{\bes}{\begin{subequations}}
\newcommand{\ees}{\end{subequations}}
\newcommand{\ord}{{\cal O}}
\newcommand{\gev}{~{\rm GeV}}
\newcommand{\tev}{~{\rm TeV}}
\newcommand{\mev}{~{\rm MeV}}
\begin{document}
\title{Petite Unification of Quarks and Leptons: Twenty Two Years After}
\author{Andrzej J. Buras}
\email[]{Andrzej.Buras@ph.tum.de}
\affiliation{Technical University Munich, Physics Department,\\
D-85748 Garching, Germany}
\author{P.Q. Hung}
\email[]{pqh@virginia.edu}
\affiliation{Dept. of Physics, University of Virginia, \\
382 McCormick Road, P. O. Box 400714, Charlottesville, Virginia 22904-4714, 
USA}
\date{\today}
\begin{abstract}
A recent surge of interest in the novel ideas of Large Extra Dimensions
and their implications, such as the {\em early} unification of quarks and
leptons, has prompted us to revive a  paper \cite{HBB} written 
twenty two years ago. In that
paper, we provided a general discussion of quark-lepton unification
characterized by the gauge group $G_{S} \otimes G_{W}$ with two
couplings $g_S$ and $g_W$ and by the unification mass scales
$M= 10\tev-1000\tev$. The constraint from $\sin^2 \theta_{W}$
restricts the choices for $G_W$ and our favorite model for
the Petite Unification (PUT) was chosen to be
$SU(4)_{\rm PS}\otimes SU(2)^4$. 
In the present paper, we review the main results of \cite{HBB} and 
propose two new models based on the groups $SU(4)_{\rm PS}\otimes SU(2)^3$ 
and $SU(4)_{\rm PS}\otimes SU(3)^2$ for which the consistency with the measured
value of $\sin^{2}\theta_{W}(M_{Z}^{2})$ determines the unification scale 
to be roughly $1~{\rm TeV}$ and $3-10~{\rm TeV}$, respectively. The 
implications of this very early unification is the existence of new quarks 
and leptons with charges up to $4/3$ (for quarks) and 2 (for leptons) 
and masses $\ord(250\gev)$. Interestingly, in these models the rare 
decay $K_L\to \mu e$ is automatically absent at tree level and the 
one-loop contributions are consistent with the experimental upper bound for
this decay. On the other hand the original $SU(4)_{PS}\otimes SU(2)^4$ 
model can only be made consistent with the measured value of 
$\sin^{2}\theta_{W}(M_{Z}^{2})$ and the unification scale 
$M=\ord (1 \tev)$, provided there exist at least {\it nine} ordinary quark and 
lepton 
generations, with {\it four} generations in the case of the supersymmetric 
version. Moreover, the solution to the $K_L\to \mu e$ problem is not as 
natural as in the two other scenarios. 
We comment on the recent papers on early unification in the context of
Large Extra Dimensions.

\end{abstract}
\maketitle

\section{Introduction}
Twenty two years ago, we have proposed alternatives to popular
Grand Unified models such as $SU(5)$ \cite{GG,BEGN} or $SO(10)$ \cite{GG1,FM}, 
based on a
less ambitious program which aimed at unifying quarks and leptons 
at some energy scale $M$ which is not too much greater than the
electroweak scale \cite{HBB}. We assumed that the Standard Model (SM),
$SU(3)_c \otimes SU(2)_L \otimes U(1)_Y$,
which has three independent couplings, $g_3$, $g_2$ and $g^\prime$, 
is embedded into a gauge theory $G_{S} \otimes G_{W}$, which is
characterized by two independent couplings $g_S$ and $g_W$,
at a ``petite unification'' scale $M$ which can be as small
as $M= 10^{5 \pm 1}\,GeV$, namely the TeV region. We further 
assumed that $G_S$ and $G_W$ are either simple or pseudosimple
(a direct product of simple groups with identical couplings). 
Our approach was a
``bottom up'' one, that is to say we used the available inputs 
from the ``low energy'' to constrain the choices of
$G_S$ and $G_W$. We used $\sin^{2} \theta_W$ and the known 
fermion representations as inputs. It turned out that the
choices of $G_W$ are quite restricted. Furthermore, if
$G_S$ is chosen to be $SU(4)$ \`{a} la Pati-Salam \cite{PASA}, 
this restriction
is even stronger, with the minimal choice for $G_W$ being
$[SU(2)]^4$ and the corresponding PUT
\be\label{PUT0}
{\rm PUT}_0=SU(4)_{\rm PS} \otimes SU(2)_{L} \otimes
SU(2)_{R} \otimes \tilde{SU(2)}_{L} \otimes \tilde{SU(2)}_{R}
\ee
This minimal model was discussed
at length in our paper. 

In the $SU(4)_{\rm PS} \otimes [SU(2)]^4$ model the value of
$\sin^{2} \theta_W$ at the unification scale $M\gg M_Z$ turns out
to be $\sin^{2} \theta_W^0=1/4$, very close to its experimental value 
that is now very precisely known: 
$\sin^{2}\theta_{W}(M_{Z}^{2})=0.23113(15)$ \cite{PDG}. 
For $M= 100\tev$ the inclusion 
of $\ord(\alpha)$ corrections and the renormalization group evolution led 
in 1981 to  $\sin^{2}\theta_{W}(M_{Z}^{2})\approx 0.22$, still consistent
with the data of 1981. As we will show below with the present value of 
the QCD coupling constant, $\alpha_s(M_Z^2)$, the consistency with 
the measured very precise value of $\sin^{2}\theta_{W}(M_{Z}^{2})$ requires 
in this model the unification scale $M$ to be as low as $330\gev$. This is 
clearly unacceptable  as the lower bound on the right-handed gauge boson 
mass is $M_{W_R}\ge 800\gev$ \cite{PDG}. 
The scale $M$ can be raised to $1\tev$ by adding six 
additional standard  fermion generations with masses $\ord(250\gev)$ or 
making the model supersymmetric, in which case two new fermion generations 
suffice. 
However in the simplest version of this model the rare decay 
$K_L\to\mu e$ proceeds at the tree level and its rate with $M=1\tev$ 
exceeds the experimental upper bound by many orders of magnitude. 
A possible solution to these difficulties, as advocated recently in 
\cite{CHHAPE}, is 
to introduce one  Large Extra Dimension to obtain acceptable values for 
$\sin^{2}\theta_{W}(M_{Z}^{2})$  and $Br(K_L\to\mu e)$ with $M=\ord(1-10)$ 
TeV and the usual 
three fermion generations. We will discuss other alternatives in this paper.

In the present paper we would like to propose two possibly more attractive 
PUT groups
\be\label{PUT1}
{\rm PUT}_1=SU(4)_{\rm PS} \otimes SU(2)_{L} \otimes
SU(2)_{H} \otimes SU(2)_{R}
\ee
and 
\be\label{PUT2}
{\rm PUT}_2=SU(4)_{\rm PS} \otimes SU(3)_L \otimes SU(3)_H
\ee
that were listed in our PUT classification of 1981, but were not analyzed 
by us in detail. In these models $\sin^{2} \theta_W^0$ equals $1/3$ and 
$3/8$, respectively but a very fast renormalization group evolution 
allows to obtain correct $\sin^{2}\theta_{W}(M_{Z}^{2})$ with 
$M=1~{\rm TeV}$ and $M=3.3~{\rm TeV}$, respectively when the spontaneous 
breakdown of the PUT groups to the Standard Model group proceeds in 
one step. Moreover, the fast renormalization 
group evolution combined with the very precise experimental value for 
$\sin^{2}\theta_{W}(M_{Z}^{2})$ determines these unification scales within 
$10-15\%$. If the breakdowns of $SU(4)_{\rm PS}$ and of $G_W$ are allowed 
to appear at two different scales $M$ and $\tilde M<M$, these two scales 
have to be close to $1\tev$ in the case of ${\rm PUT}_1$ but can differ up 
to an order of magnitude in the case of ${\rm PUT}_2$ with roughly 
$3\le M\le 10\tev$ and $0.8\le \tilde M\le 3\tev$.

These two scenarios for early unification of quark and leptons have three 
interesting properties:
\begin{itemize}
\item
In addition to the standard three generations of quarks and leptons, new 
three generations of unconventional quarks and leptons
with charges up to $4/3$ (for quarks) and 2 (for leptons) 
and masses $\ord(250\gev)$ are automatically present.  
The horizontal groups $SU(2)_H$ and $SU(3)_H$ connect the standard
fermions with the unconventional ones.
\item
The placement of the ordinary quarks and leptons in the fundamental 
representation 
of $SU(4)_{\rm PS}$ is such that there
are {\em no tree-level} transitions between ordinary quarks and leptons
mediated by the $SU(4)_{\rm PS}$ gauge bosons. This prevents rare decays 
such as $K_L \rightarrow \mu e$ from acquiring
large rates, even when the masses of these gauge bosons are 
in the few TeV's range.
\item
There are new contributions to flavour changing neutral current
processes (FCNC) involving standard quarks and leptons that are mediated 
by the horizontal $SU(2)_H$ and $SU(3)_H$ weak gauge bosons and the new
unconventional quarks and leptons. However, they appear first at the 
one--loop level and can be made consistent with the existing experimental 
bounds. 
\end{itemize}

Our paper is organized as follows. In Sec. II, we review 
the steps that lead to the three choices for $G_W$ mentioned above and 
we summarize the most important formulae. In particular we derive 
the general expression for $\sin^{2} \theta_W^0$ and discuss its relation to
$\sin^{2}\theta_{W}(M_{Z}^{2})$. In section III we present in detail 
the fermion content of the selected groups. The results of the 
renormalization group
analysis of $\sin^{2} \theta_W$, in the scenarios in question, is presented
in Sec. IV and in Sect. V the rare decay $K_L\to \mu e$ is briefly
discussed.
Here we emphasize that while in the $SU(4)_{\rm PS} \otimes SU(2)^4$ 
scenario, it is very difficult to satisfy the experimental bound 
on $K_L\to \mu e$ when $M=\ord(1\tev)$, the presence of GIM--like 
mechanism in the remaining two scenarios allows to satisfy this bound 
without any unnatural conditions on the mass spectrum of new quarks 
and leptons and related CKM-like mixing matrix. Similar comments 
apply to FCNC processes.

  In Sec. VI we compare our work of 1981 and the one presented 
here with the recent papers on the early unification of quarks and leptons 
in the context of Large Extra Dimensions \cite{CHHAPE,DIKA}. 
As a matter of fact the $SU(3)_W$ model of Dimopoulos and Kaplan 
\cite{DIKA} is just one of the cases considered by us in \cite{HBB} 
and the analysis in \cite{CHHAPE} is the generalization of our 
$SU(4)_{\rm PS} \otimes [SU(2)]^4$ model to extra dimensions.
Finally, in Sec. VII we summarize the main results of our paper and 
offer some perspectives for the future work. Detailed analysis of 
$K_L\to \mu e$ and other phenomenological implications of the 
PUT groups discussed here will be presented elsewhere. 

\section{Petite Unification revisited}
\subsection{Preliminaries}
The objective, then and now, is to unify quarks and leptons
at an intermediate scale in the TeV range. We assume, then
and now, that $SU(3) \otimes SU(2)_L \otimes U(1)_Y$ is
embedded in $G = G_{S}(g_S) \otimes G_{W}(g_W)$, where
$g_S$ and $g_W$ denote the corresponding couplings. Furthermmore,
$G_S$ and $G_W$ are assumed to be either simple or pseudosimple, i.e.,
a direct product of simple groups with identical couplings.
The pattern of symmetry breaking is assumed to be
\begin{equation}
\label{pattern}
G \stackrel{\textstyle M}{\longrightarrow} G_1 
\stackrel{\textstyle \tilde{M}}{\longrightarrow} G_2
\stackrel{\textstyle M_Z}{\longrightarrow} SU(3)_c \otimes U(1)_{EM} ,
\end{equation}
where
\begin{equation}
\label{G1}
G_1 = SU(3)_{c}(g_3) \otimes \tilde{G}_{S}(\tilde{g}_S) \otimes
G_{W}(g_W) \, ,
\end {equation}
and
\begin{equation}
\label{G2}
G_2 = SU(3)_{c}(g_3) \otimes SU(2)_{L}(g_2) \otimes U(1)_{Y}(g^\prime)\,.
\end {equation}
We assume $M_Z < \tilde{M} \leq M$. In principle, $G$ can be
broken down directly to $G_2$, but to be more general, the
pattern (\ref{pattern}) was assumed in \cite{HBB}.
Furthermore, in accordance with our petite-unification idea,
we require 
\begin{itemize}
\item
 $M$ and $\tilde{M}$ to be at most  a few orders of
magnitude larger than $M_Z$, 
\item
the weak hypercharge $U(1)_{Y}$
group to merge into both $\tilde{G}_S$ and $G_W$ at $\tilde{M}$,
\item
$SU(3)_c$ and $\tilde{G}_S$ to be unbroken subgroups
of $G_S$ so that their generators are unbroken generators of
$G_S$.
\end{itemize}
The second requirement  allows us to put quarks and leptons into identical
representations of the weak group $G_W$ and consequently make the quarks
 and leptons to be indistinguishable when the strong interactions are turned 
off. The last requirement implies 
\be
g_{3}(M^2)= \tilde{g}_{S}(M^2) = g_{S}(M^2) \, .
\ee

\subsection{\boldmath{$\sin^{2}\theta_{W}$} and the choices of $G_{W}$}

We will next summarize the salient points of our earlier
paper concerning the restrictions imposed on $G_W$ from the value
of $\sin^{2}\theta_{W}$. We will focus, in particular, on the case
where $G_{W} = [SU(N)]^k$ and use $\sin^{2}\theta_{W}$ to constrain
the pair $(N,k)$. Furthermore,
we have argued in \cite{HBB} that the most economical choice
for $G_S$ is $SU(4)$ \`{a} la Pati-Salam although we have
presented there a more general discussion. In the following
we shall then deal principally with the groups 
\be\label{GPUT}
G=SU(4)_{\rm PS} \otimes [SU(N)]^k, \qquad \tilde{G}_S= {U(1)}_{\rm S}. 
\ee
To derive $\sin^{2} \theta_W$, we write the
generators $T_{3L}$ and $T_0$ of $SU(2)_L$ and $U(1)_Y$
respectively, in terms of the generators of $G_S$ and 
$G_W$. As usual, one has for the electric charge generator $Q$
\begin{equation}
\label{Q00}
Q = T_{3L} + T_{0}\,,
\end{equation}
where $T_{3L}$ and $T_{0}$ are diagonal generators of $SU(2)_L$
and $U(1)_Y$, respectively. They can be written as
\begin{equation}
\label{T3L}
T_{3L} = \sum_{\alpha} C^{'}_{\alpha W} T^{0}_{\alpha W}\, ,
\end{equation}
and
\begin{equation}
\label{T0}
T_{0} = \sum_{\alpha} C_{\alpha W} T^{0}_{\alpha W} +
C_{S} T_{15} \, ,
\end{equation}
where $T^{0}_{\alpha W}$ and $T_{15}$ are
the diagonal generators of $G_W$ and $SU(4)_{PS}$ respectively,
with $T^{0}_{\alpha W}$ being the generators of the $SU(2)$
disjoint subgroups of $G_W$.
Also, $C^{'}_{\alpha W}$ and $C_{\alpha W}$ are orthogonal
to each other. 

Eqs.(\ref{T3L},\ref{T0}) form the basis for the derivation
of $\sin^{2}\theta_{W}$. In \cite{HBB}, we discussed two
cases which were called (a) the ``unlocked standard model''
where the generators of $SU(2)_L$ are the unbroken generators
of $G_W$, and (b) the ``locked standard model'' where
the generators of $SU(2)_L$ are the unbroken combination of
generators belonging to several disjoint $SU(2)$ subgroups
of $G_W$. We showed that case (a) (the ``unlocked standard model'')
is the most economical one and this is one we will choose
to concentrate on in the present paper. The reader is encouraged
to consult \cite{HBB} for a more general discussion.
Therefore for case (a), one
has
\begin{equation}
\label{T3Lp}
T_{3L} = T^{0}_{3W} \,,
\end{equation}
where $T^{0}_{3W}$ is a diagonal generator of one of $SU(2)$ 
subgroups of $G_W$. This implies that  $C^{'}_{3 W} =1$ with all
other coefficients in (\ref{T3L}) equal to zero.
In consequence, in the ``unlocked standard model'' scenario,
one is now in a position to derive $\sin^{2}\theta_{W}$, taking
into account the pattern (\ref{pattern}). First, we present
a formula for the renormalized value of $\sin^{2}\theta_{W}$
at the one-loop level. We will comment on its generalization to
two loops in Sec. IV. From
\begin{equation}
\label{e2}
\frac{1}{e^{2}(M_{Z}^{2})} =\frac{1}{[g_{2}(M_{Z}^{2})]^2} +
\frac{1}{[g^{'}(M_{Z}^{2})]^2}\, ,
\end{equation}
\begin{equation}
\label{g2}
g_{2}(\tilde{M}^2)= g_{W}(\tilde{M}^2)\, ,
\end{equation}
\begin{equation}
\label{e2p}
\frac{1}{[g^{'}(\tilde{M}^{2})]^2} =\frac{\sum_{\alpha}C_{\alpha W}^2}
{[g_{W}(\tilde{M}^{2})]^2} +\frac{C_{S}^2}
{[\tilde{g}_{S}(\tilde{M}^{2})]^2}\, ,
\end{equation}
and using the $\overline{MS} $ definition for $\sin^{2}\theta_{W}$, namely
\begin{equation}
\label{sinsq}
\sin^{2}\theta_{W}(M_{Z}^{2}) = \frac{e^{2}(M_{Z}^{2})}
{g_{2}^{2}(M_{Z}^{2})}\, ,
\end{equation}
one obtains  the master formula \cite{HBB}
\begin{eqnarray}
\label{sinsq2}
\sin^{2}\theta_{W}(M_{Z}^{2})& =& \sin^{2}\theta_{W}^{0}[1-
C_{S}^{2}\frac{\alpha(M_{Z}^{2})}{\alpha_{S}(M_{Z}^{2})}-8\pi\times 
\nonumber \\
& &  \alpha(M_{Z}^{2})(K\ln\frac{\tilde{M}}{M_Z}+
K^{'}\ln\frac{{M}}{\tilde M})]\, ,
\end{eqnarray}
where
\be
\alpha(M_{Z}^{2}) \equiv \frac{e^{2}(M_{Z}^{2})}{4\pi}, \qquad
\alpha_{S}(M_{Z}^{2}) \equiv \frac{g_{3}^{2}(M_{Z}^{2})}{4\pi},
\ee
\begin{equation}
\label{sinsq0}
\sin^{2}\theta_{W}^{0} = \frac{1}{1+C_{W}^2} \,,
\end{equation}
with $C_{W}^2 = \sum_{\alpha}C_{\alpha W}^2$, and
\begin{equation}
\label{K}
K = b_1 - C_{W}^{2}b_2 - C_{S}^{2} b_3 \, ,
\end{equation}
\begin{equation}
\label{K'}
K^{'} = C_{S}^2 (\tilde{b} - \tilde b_3) \, .
\end{equation}
Here, $b_1$, $b_2$, $b_3 (\tilde b_3)$, and $\tilde{b}$ are the one-loop
coefficients of the beta functions for $U(1)_Y$, $SU(2)_L$,
$SU(3)_c$, and $U(1)_S$, respectively  with $\tilde b_3\not=b_3$ due to 
possible contributions of new particles with masses larger than $\tilde M$. 
Explicit expressions for these coefficients are given in section IV.
We will see there that in the case of the new groups in (\ref{PUT1}) and
(\ref{PUT2}), the presence of new particles with masses $\ord{(250\gev)}$
will require the introduction of the appropriate threshold corrections in
$K$.

Neglecting the contributions of new particles to $K$ and $K^\prime$ for a
moment and using the $\overline{\rm MS}$ values \cite{PDG}
\begin{equation}\label{newin1}
1/\alpha(M^2_Z)= 127.934(27), \quad 
\alpha_S(M^2_Z)= 0.1172(20)
\end{equation}
we find
\be
\sin^{2}\theta_{W}(M_{Z}^{2})= R \sin^{2}\theta_{W}^{0}
\ee
where
\begin{eqnarray}
\label{sinsq3}
R & =& 1-0.067 C_{S}^{2}-0.014 C_{S}^{2}\ln\frac{{M}}{\tilde M}   
\nonumber \\
& & -( 0.009 + 0.004 C^2_W + 0.009 C^2_S)\ln\frac{\tilde{M}}{M_Z}\,.
\end{eqnarray}
We observe that $\sin^{2}\theta_{W}(M_{Z}^{2})$ is a sensitive function 
of $C^2_W$, present in particular in $\sin^{2}\theta_{W}^{0}$, and of 
$C^2_S$ in the term  
$C_{S}^{2}\alpha(M_{Z}^{2})/\alpha_{S}(M_{Z}^{2})$ and in the renormalization
group corrections. The renormalization of $\sin^{2}\theta_{W}$ 
increases with increasing  $C^2_S$ but of course depends also strongly on 
the values of $b_i's$, that in turn depend on the content of
the fermion representations and their weak and strong charges.  

As $\sin^{2}\theta_{W}(M_{Z}^{2})$ is known with a very 
high precision,
\begin{equation}\label{newin2}
\sin^{2}\theta_{W}(M_Z^2)|_{\rm exp}= 0.23113(15)~, 
\end{equation}
and $C_W$ and $C_S$ in the case of $SU(4) \otimes [SU(N)]^k$
can take only special values, only certain pairs $(C_W,C_S)$
are allowed if we are interested in the unification scales
$M\le 1000\, TeV$ and in particular $M\le 10\, TeV$.
We will now briefly describe the steps that led us in \cite{HBB} to the
acceptable choices of $(C_W,C_S)$.

The crucial quantity to be considered first is
$\sin^{2}\theta_{W}^{0}$ which is determined at the petite
unification scale $M$. For $G_{W} = [SU(N)]^k$, a given
pair of $(N,k)$ will determine $C_W$ and hence
$\sin^{2}\theta_{W}^{0}$ through Eq. (\ref{sinsq0})
which can also be written as
\begin{equation}
\label{sinsq02}
\sin^{2}\theta_{W}^{0} = \frac{1}{1+C_{W}^2} =
[\frac{TrT_{3L}^2}{TrQ^2}]_{adjoint} \,,
\end{equation}
where the last term in (\ref{sinsq02}) reflects the fact
that the adjoint representation of $G_W$ is a singlet
of $G_S$. It is then sufficient to evaluate (\ref{sinsq02})
by simply examining the adjoint representation.

Since quarks and leptons are assumed to be in separate (but identical)
representations of $G_W$, the gauge bosons of $G_W$ have integer
charges. Assuming next a permutation symmetry among the $SU(N)$'s in $G_W$, 
and allowing for arbitrary integer charges for the
gauge bosons one finds \cite{HBB}
\begin{equation}
\label{sinsq03}
\sin^{2}\theta_{W}^{0} = \frac{N}{k Tr(Q_{W}^2)|_{adj}}, \qquad
Tr(Q_{W}^2)|_{adj} = \sum_{i=1}^{\alpha} i^2 n_{i} \, ,
\end{equation}
 where $Tr(Q_{W}^2)|_{adj}$ is for each $SU(N)$, 
$n_{i}$ is the number of gauge bosons with $|Q| =i$,
and $\alpha$ is the maximal gauge-boson charge involved. Since
the adjoint representation can be constructed from the product
of the fundamental representation $N$ and its conjugate $\bar{N}$,
one can compute $n_i$ by looking at the charge distribution of
the fundamental representation, namely
\begin{equation}
\label{fundmental}
[\underbrace{\tilde{Q}_{W},\cdots \tilde{Q}_{W}}_{r_0},
\underbrace{\tilde{Q}_{W}-1,\cdots \tilde{Q}_{W}-1}_{r_1},
\underbrace{\tilde{Q}_{W}-\alpha,\cdots \tilde{Q}_{W}-\alpha}_{r_\alpha}]\, ,
\end{equation}
where $\tilde{Q}_W$ is an eigenvalue of $Q_W$.

The detailed analysis in \cite{HBB} has shown that
\begin{itemize}
\item
Gauge bosons with charges $\pm 3$ or higher 
corresponding to $N\geq 4$ are excluded
since one can derive the inequality $\sin^{2}\theta_{W}^{0} \leq
1/(12-(8/N)) \leq 1/10$ which rules out this case.
\item
For doubly charged gauge bosons, the maximal allowed number
is two (for $\pm2$) leading to $Tr(Q_{W}^2)=4N$ for any $SU(N)$
with $N \geq 3$. For $k=1$ this gives $\sin^{2}\theta_{W}^{0}=1/4$.
However, as shown in \cite{HBB}, in this case $C^2_S=8/3$, implying 
through (\ref{sinsq3}) $\sin^{2}\theta_{W}(M_{Z}^{2})=0.205$
even without including the renormalization group effects that decrease 
it even further. As a consequence, scenarios with
$G_W = SU(3),\, SU(4), \ldots,$ having two doubly charged gauge bosons
are inconsistent with the data.
\end{itemize}

We thus obtain an important result: 
\begin{itemize}
\item
the only charges of weak gauge bosons 
that are consistent with the measured value of  
$\sin^{2}\theta_{W}(M_{Z}^{2})$ within the petite unification framework 
with the gauge group $SU(4) \otimes [SU(N)]^k$ are $0$ and $\pm1$.
\end{itemize}
Consequently the formula (\ref{sinsq03}) simplifies to 
\begin{equation}
\label{simple}
\sin^{2}\theta_{W}^{0} = \frac{N}{k n_1}, 
\end{equation}
where $n_1$ is the number of weak gauge bosons with $Q=\pm 1$ in $SU(N)$.

In order to find $n_1$ let us consider first the class (i) of fermion 
representations that transform under $G_W$ as
\begin{equation}
\label{rep}
(f,1,1 \cdots, 1),\, (1,f,1,\cdots,1) \,.
\end{equation}
Each entry in (\ref{rep}) corresponds to the group $\tilde G$ in the 
product $G_W=\tilde G\otimes \tilde G\cdots \otimes \tilde{G}$. 
That is quarks and leptons 
transform nontrivially under one of the groups $\tilde G$ and are singlets 
under the rest. The fundamental representation for the group $\tilde{G}$
has then a charge distribution
\begin{equation}
\label{fundamental2}
[\underbrace{\tilde{Q}_{W},\cdots \tilde{Q}_{W}}_{r_0},
\underbrace{\tilde{Q}_{W}-1,\cdots \tilde{Q}_{W}-1}_{r_1}]\, ,
\end{equation}
with $r_0 + r_1 = N$. The tracelessness condition for the charge
operator $Q_W$ gives the eigenvalues
\begin{equation}
\label{eigenvalue}
\tilde{Q}_{W} = 1 -\frac{r_{0}}{N},\,\quad \tilde{Q}_{W} -1 \, .
\end{equation}
Moreover we find 
\be
n_1=2 r_0 r_1=2 r_0(N-r_0)
\ee
 and consequently  a very useful formula
\begin{equation}
\label{simple2}
\sin^{2}\theta_{W}^{0} = \frac{N}{2k r_0 (N-r_0)}=\frac{1}{1+C_{W}^2}, 
\end{equation}
that can be used to calculate $\sin^{2}\theta_{W}^{0}$ and 
$C_W^2$ for given $N$, $k$ and $r_0$. 
This formula is equivalent to the formulae given in \cite{HBB} but is 
more transparent.
The results for $\sin^{2}\theta_{W}^{0}$ are given in 
table~\ref{tab:table1}, where also the values of the charges $\tilde Q^i_W$
in the fundamental representation obtained by means of (\ref{eigenvalue}) 
are given. We observe a correlation between the values of 
$\sin^{2}\theta_{W}^{0}$ for given $(N,k)$ and the weak charges of quarks
and leptons. This correlation implies eventually the correlation between
$\sin^{2}\theta_{W}^{0}$ and electric charges of quarks and leptons that 
follows from 
\begin{equation}
\label{Q}
Q=Q_S+Q_W=C_S T_{15}+Q_W
\end{equation}
where $T_{15}$ is the diagonal generator of $SU(4)_{\rm PS}$ that commutes 
with $SU(3)_c$. We will return to this correlation below.

If fermions transform as (class(ii)) $(f,\bar f)$ under any pair 
$\tilde G\otimes \tilde G$ in $G_W$ and are singlets under the rest, 
that is in the symbolical notation of (\ref{rep}) one has
\begin{equation}
\label{rep22}
(f,\bar{f}, 1,\cdots,1) \, ,
\end{equation}
the charge distribution is a $N \times N$ matrix with
$r_{0}+r_{1}=N$ columns and $r_{0}^{'}+r_{1}^{'}=N$ rows
(see Eq. (4.10) of \cite{HBB}).  
This matrix looks like:
\begin{equation}
\left(
\begin{array}{cccccc}
\tilde{Q}_{W}&\ldots & \tilde{Q}_{W}&\tilde{Q}_{W}-1&\ldots&\tilde{Q}_{W}-1 \\ 
& \vdots & & \\
\tilde{Q}_{W}&\ldots & \tilde{Q}_{W}&\tilde{Q}_{W}-1&\ldots&\tilde{Q}_{W}-1 \\
\tilde{Q}_{W}+1&\ldots & \tilde{Q}_{W}+1&\tilde{Q}_{W}&\ldots&\tilde{Q}_{W} \\ 
& \vdots & & \\
\tilde{Q}_{W}+1&\ldots & \tilde{Q}_{W}+1&\tilde{Q}_{W}&\ldots&\tilde{Q}_{W}
\end{array}
\right)\ \,,
\label{chargedis}
\end{equation}
where the rows refer to $f$ and the columns to $\bar{f}$.
The eigenvalues of $Q_W$
are now \cite{HBB}
\begin{equation}
\label{eigenvalue2}
\tilde{Q}_{W} = \frac{r_{0}^{'}-r_{0}}{N},\, \quad \tilde{Q}_{W}\pm 1 \, .
\end{equation}

It turns out that from the point of view of $\sin^{2}\theta_{W}$ only 
the cases $r^\prime_0=r_0$ and consequently 
$r^\prime_1=r_1=N-r_0$ are of interest to us implying $\tilde Q_W=0,\pm 1$
as shown in table~\ref{tab:table1}. Moreover the formula (\ref{simple2}) 
also applies here.

Whether the groups listed in table~\ref{tab:table1} give the acceptable 
$\sin^{2}\theta_{W}(M^2_Z)$ depends also on $C^2_S$ as discussed before. 
In fact it has been shown in \cite{HBB} that if 
$G_S$ was chosen to be the Pati-Salam $SU(4)$ with each standard quark 
$SU(3)_c$ triplet put with a lepton  into the same fundamental representation
of $SU(4)$ and the electric charges of quarks and leptons are restricted to
\begin{equation}\label{charges}
Q_q=\frac{d}{3}+n, \qquad Q_l=n^\prime, \qquad n,n^\prime~{\rm integer}, 
\quad d=1,2~,
\end{equation}
then many of the possibilities given in table~\ref{tab:table1} can be 
eliminated. The choice in (\ref{charges}) allows to include at least quarks 
and leptons with ordinary charges. Indeed under the latter assumption one can 
show that $\tilde{Q}_{W}^{i}$
should be multiples of $1/4$, in fact
\be
\label{Qi}
\tilde{Q}_{W}^{i} = \frac{1}{4}(3Q_{q}^{i} + Q_{l}^{i}) \, .
\ee
Consequently a number of possibilities listed in table~\ref{tab:table1} can be 
eliminated only by this requirement. For the remaining cases that 
satisfy (\ref{Qi}) we find using
\be\label{ch}
Q^i_q=\frac{C_S}{2\sqrt{6}} + \tilde Q^i_W~, \quad 
Q^i_l=-\frac{3C_S}{2\sqrt{6}} + \tilde Q^i_W~,
\ee
the expression for $C^2_S$ in terms of quark and lepton electric charges
\begin{equation}
\label{CS}
C_{S}^{2} = \frac{1}{6}(3Q_{q}^{i}-3Q_{l}^{i})^2 .
\end{equation}

One word of caution is in order here. The previous statements related
to (\ref{charges}) refer only to scenarios in which the only
representations present are of a single class, i.e. (i) or (ii). In the
case where both classes are needed, as  will be 
the case of ${\rm PUT}_1$,
we should broaden the restriction (\ref{charges})
in the following sense. First, the value of $C_{S}^{2}$ should be
chosen judiciously depending on $\sin^{2}\theta_{W}^{0}$. Once it
is chosen, the charges of the fermions are determined depending on
their representations under $G_W$ and are given by Eq. (\ref{Q}),
namely $Q=C_S T_{15}+Q_W$. As we have discussed earlier and shown in
Table 1, representations $(f,1,1,..)$ have $Q_W = \pm 1/2$ and
representations $(f,\bar{f},1,...)$ have $Q_W = 0, \pm 1$. Obviously,
when a scenario contains both classes of representations, it will
be unavoidable to have quarks and leptons with ``funny'' charges
in addition to the familiar ones. As we will discuss below in the context 
of ${\rm PUT}_1$,
 as long as some of  these ``funny'' fermions belong
to a vector-like representation of one of the $G_W$ gauge groups,
they can be very massive, in the sense that their masses are not
proportional to the SM electroweak scale. The obvious caution
that one has to take is that, in a mixed case, at least one of the
representations has to contain SM fermions.

\begin{table}
\caption{\label{tab:table1}. The values of $\sin^2\theta^0$ for the 
weak groups $G_W=SU(N)^k$ and different fermion representations.}
\begin{ruledtabular}
\begin{tabular}{ccccc} 
 & & &($f$,1)+(1,$\bar{f}$)& ($f$,$\bar{f}$) \\
$G_W$ & $r_0$ & $\sin^{2}\theta_{W}^{0}$ & $\tilde{Q}_{W}^{i}$ 
& $\tilde{Q}_{W}^{i}$ \\ \hline
$[SU(2)]^3$ & 1 & 0.333 & $\pm\frac{1}{2}$ & 0,$\pm1$ \\
$[SU(2)]^4$ & 1 & 0.250 & $\pm\frac{1}{2}$ & 0,$\pm1$ \\
$[SU(3)]^2$ & 1 & 0.375 & $\frac{2}{3}$,$-\frac{1}{3}$ & 0,$\pm1$ \\
$[SU(3)]^3$ & 1 & 0.250 & $\frac{2}{3},-\frac{1}{3}$ & 0,$\pm1$ \\
$[SU(4)]^2$ & 2 & 0.250 & $\pm\frac{1}{2}$ & 0,$\pm1$ \\
$[SU(5)]^2$ & 1 & 0.313 & $\frac{4}{5}$,$-\frac{1}{5}$ & 0,$\pm1$ \\
$[SU(6)]^2$ & 1 & 0.300 & $\frac{5}{6},-\frac{1}{6}$ & 0,$\pm1$ \\
$SU(7)$ & 3 & 0.292 & $\frac{4}{7},-\frac{3}{7}$&   \\
$[SU(7)]^2$ & 1 & 0.292 & $\frac{6}{7},-\frac{1}{7}$ & 0,$\pm1$ \\ 
$SU(8)$ & 3 & 0.267 & $\frac{5}{8},-\frac{3}{8}$&   \\
$SU(8)$ & 4 & 0.250 & $\pm\frac{1}{2}$&   \\
\end{tabular}
\end{ruledtabular}
\end{table}

With the condition on $Q^i_{q,l}$ in (\ref{charges}) the lowest values for 
$C^2_S$ are found to be
\be
C^2_S=\frac{1}{6},~\frac{2}{3},~\frac{8}{3}.
\ee
The next value $C^2_S=25/6$ and higher values would require very small
$\sin^{2}\theta_{W}(M^2_Z)$ and rather high quark and lepton charges.
In table~\ref{tab:table2} we list $\tilde{Q}_{W}^{i}$ of 
table~\ref{tab:table1} 
which satisfies Eq. (\ref{Qi}) along with the corresponding quark and
lepton charges, as well as the values of $C_{S}^2$. 
Although, for completeness, we also list the case $C_{S}^{2} =
1/6$ in table II, it has been shown in \cite{HBB} that it
corresponds to a weak group $SU(4)_{1} \otimes SU(4)_{2}$ which
has $\sin^{2}\theta_{W}^{0}=0.286$. Because of the low value of $C^2_S$, 
one needs $M > 10^6\gev$ in order to obtain the correct value of 
$\sin^{2}\theta_{W}(M_Z^2)$ and consequently this scenario does not fit into 
our framework.

We now classify the $G_W$ groups listed in table~\ref{tab:table1}  in terms of
their possible agreements with $\sin^{2}\theta_{W}(M_Z^2)$.
As seen from tables~\ref{tab:table1} and \ref{tab:table2} only the values
$C^2_S=2/3,8/3$ have to be considered.
We can make then the following observations.

(a) Groups which can have $C_{S}^2 = 2/3$ are those for which
$\tilde{Q}_{W}^i = \pm 1/2$ which corresponds to representations
which contain only conventionally charged quarks and leptons, as can be
seen from Table~\ref{tab:table2}. From Table~\ref{tab:table1}, these 
weak groups 
are
$[SU(2)]^3$, $[SU(2)]^4$,
$[SU(4)]^2$ and $SU(8)$, 
with $\sin^{2}\theta_{W}^{0} = 0.333,
0.25,0.25,0.25$, respectively. For $[SU(2)]^3$, one
would  need a petite unification scale substantially larger than
1000 TeV because $C_{S}^2 = 2/3$ is too small to bring
$\sin^{2}\theta_{W}^{0} = 0.333$ down to 
$\sin^{2}\theta_{W}(M_Z^2) \sim 0.23$. (We shall however come back
to this group in the discussion below.) The promising groups
in this class of models are, in order of complexity,
$[SU(2)]^4$, $[SU(4)]^2$ and 
$SU(8)$,
all of which have $\sin^{2}\theta_{W}^{0} = 0.25$. In particular,
the group $[SU(2)]^4$ was our favorite choice in \cite{HBB}.
The renormalization
group (RG) analysis of these models will be discussed in Sec. IV.

(b) Groups that have $C_{S}^2 = 8/3$ are those with
$\tilde{Q}_{W}^i = 0, \pm 1$ which corresponds to representations
having quark charges as high as $\pm 4/3$ and lepton charges
as high as $\pm 2$ in addition to the standard charges. 
Because of the high value for $C_{S}^2$,
we need those groups for which $\sin^{2}\theta_{W}^{0} > 0.3$.
From Table ~\ref{tab:table2},
one can see that only three groups satisfy this criterion:
$[SU(2)]^3$, $ [SU(3)]^2$,
and $[SU(5)]^2$, with 
$\sin^{2}\theta_{W}^{0} = 0.333, 0.375, 0.313$, respectively.
The implications of the first two of these models through a RG analysis will
be discussed in Sec. IV.

\begin{table}
\caption{\label{tab:table2}. The values of lepton ($Q^i_l$) and quark 
($Q^i_q$) electric charges corresponding to the weak charges 
($\tilde Q^i_W$) discussed in the text. The values of $C^2_S$ have been 
obtained from (\ref{CS}).}
\begin{ruledtabular}
\begin{tabular}{cccc}
$\tilde{Q}_{W}^{i}$& $Q_{l}^{i}$&$Q_{q}^{i}$&$C_{S}^2$ \\ \hline
$\frac{1}{2}$& 0& $\frac{2}{3}$& \\
$-\frac{1}{2}$&-1&$-\frac{1}{3}$& \\
$\frac{1}{2}$&1&$\frac{1}{3}$&$\frac{2}{3}$ \\
$-\frac{1}{2}$& 0& $-\frac{2}{3}$& \\ \hline
1&0&$\frac{4}{3}$& \\
0&-1&$\frac{1}{3}$& \\
-1&-2&$-\frac{2}{3}$&$\frac{8}{3}$ \\
1&2&$\frac{2}{3}$& \\
0&1&$-\frac{1}{3}$& \\
-1&0&$-\frac{4}{3}$& \\ \hline
$\frac{5}{4}$& 1& $\frac{4}{3}$& \\
$\frac{1}{4}$& 0& $\frac{1}{3}$&$\frac{1}{6}$ \\
$-\frac{3}{4}$& -1& $-\frac{2}{3}$& \\
\end{tabular}
\end{ruledtabular}
\end{table}

In summary, we have arrived at two classes of weak gauge groups $G_W$
which with $G_S=SU(4)_{\rm PS}$ might satisfy the experimental constraint
on $\sin^{2}\theta_{W}(M_Z^2)$: 
\begin{itemize}
\item
\be\label{group1}
[SU(2)]^4,\quad [SU(4)]^2,\quad SU(8),
\ee
which have only conventionally charged quarks and leptons 
in the fundamental representations in (\ref{rep}), $C^2_S=2/3$
and
$\sin^{2}\theta_{W}^{0}=0.25$.
\item
\be\label{group2}
[SU(2)]^3,\quad  [SU(3)]^2, \quad [SU(5)]^2, 
\ee
which contain extra quarks and leptons with higher charges
($\pm 4/3$ and $\pm 2$) placed together with the standard quark and leptons 
in the representations (\ref{rep22}). See also Table~\ref{tab:table2}.
These groups have  respectively higher initial
$\sin^{2}\theta_{W}^{0} = 0.333,~ 0.375,~ 0.313$ and $C^2_S=8/3$.
\end{itemize}

\section{Fermion content of selected groups}
\label{content}
\subsection{Preliminaries} 
In this section we will present in detail the fermion content of three
groups, ${\rm PUT_0}$, ${\rm PUT_1}$ and ${\rm PUT_2}$ as defined in 
(\ref{PUT0}), (\ref{PUT1}) and (\ref{PUT2}), respectively.  
As we shall see in the next section,
these three groups seem to be the best candidates for a successful
Petite Unification consistent with the measured value of 
$\sin^{2}\theta_{W}$. 
The values for
$\sin^{2}\theta_{W}^{0}$ in these three scenarios are $1/4$, $1/3$ and 
$3/8$, respectively with the
latter being very reminescent of the quintessential $SU(5)$ value. Our 
analysis of the previous section implies then that the only chance to 
satisfy the $\sin^{2}\theta_{W}$ constraint is to choose for these three 
groups $C_{S}^{2}$ equal to $2/3$, $8/3$ and $8/3$, respectively. 
In other words, as one can deduce from Table 2, we should have class (i)
representation i.e. $(4,2,1,1,1),\, (4,1,2,1,1),\,(4,1,1,2,1),
(4,1,1,1,2)$ for $SU(4)_{S} \otimes [SU(2)]^4$, and 
class (ii) representation i.e. $(4, 3,\bar{3})$ for 
$SU(4)_{S} \otimes [SU(3)]^2$ .
On the other hand we will show that, for $SU(4)_{S} \otimes [SU(2)]^3$,
both classes are involved.

While the value of $C^2_S$ is an important ingredient in the relation 
between $\sin^{2}\theta_{W}^0$ and $\sin^{2}\theta_{W}(M^2_Z)$, the values 
of the renormalization group coefficients $b_i$ that enter $K$ and $K'$
in (\ref{K}) and (\ref{K'}) are equally important. 
In order to find these values in the scenarios considered, it is necessary 
to identify the fermion representations and the relevant charges with 
respect to the SM group and $U(1)_S$. This is what we intend to do next.

\subsection{\boldmath{ $SU(4)_{\rm PS} \otimes [SU(2)]^4$}}

This scenario has been already worked out in detail in \cite{HBB} 
and we will only recall the most important points. The weak group
\be
[G_W]_0= 
SU(2)_{L} \otimes SU(2)_{R} \otimes \tilde{SU(2)}_{L} 
\otimes \tilde{SU(2)}_{R}
\ee
consists of the standard weak gauge group of the Pati-Salam model 
and its ``mirror group'' $\tilde{SU(2)}_{L}\otimes \tilde{SU(2)}_{R}$
necessary to obtain the correct $\sin^2\theta_W$. In the original 
Pati-Salam model \cite{PASA} one has $\sin^2\theta_W^0=1/2$, 
that is much too high 
for an early unification with $C_S^2=2/3$. 
We will return to it in section IV.

Let us denote by $l_L$, the usual left-handed lepton $SU(2)_L$
doublet, and by $q_L$ the left-handed quark doublet.
The $SU(2)_R$ doublets are denoted by $l_R$ and $q_R$. Similarly,
the $\tilde{SU(2)}_{L,R}$ doublets will be denoted by
$\tilde{l}_{L,R}$ and $\tilde{q}_{L,R}$. Consequently, each 
generation of $SU(4)_{PS} \otimes [SU(2)]^4$  can be written as
\be
\Psi_{L} = (q_L, l_L)=(4,2,1,1,1)_L~,
\ee
\be
\Psi_{R} = (q_R, l_R)=(4,1,2,1,1)_R~,
\ee        
\be
\tilde{\Psi}_{L} = (\tilde{q}_L, \tilde{l}_L)
=(4,1,1,2,1)_L~,
\ee
\be
\tilde{\Psi}_{R} = (\tilde{q}_R, \tilde{l}_R)=
(4,1,1,1,2)_R~.
\ee
$\tilde{\Psi}_{L}$ and $\tilde{\Psi}_{R}$ are what we call
``mirror fermions''.

Note that in this scenario the weak charges in each $SU(2)$ representation
are
\be\label{f11}
Q_W=(1/2,-1/2)
\ee
and with $C^2_S=2/3$,
\be
Q_q^i=\frac{1}{6}+Q^i_W, \qquad Q_l^i=-\frac{1}{2}+Q^i_W~.
\ee
Consequently only conventional electric charges are present and they are the 
same for the ordinary and mirror fermions. However, the latter are 
 $SU(2)_L$ (as well as $SU(2)_R$) singlets.

Now, in order to have a ``Petite Unification''
with only two independent couplings, $g_S$ and $g_W$,  
the four gauge couplings of $[SU(2)]^4$ have to be equal to each other 
above the scale $\tilde{M}$. Consequently the 
mirror fermions have to be {\em lighter} than $\tilde{M}$.
Below $\tilde{M}$, the 
masses of mirror fermions and possible extra generations are
however unconstrained, although the detailed spectrum depends on the 
Higgs system used to generate the fermion masses.
As discussed in \cite{HBB}, the appropriate Higgs scalars
which could give masses to the normal and mirror fermions can transform as
$(1,2,2,1,1)$ and $(1,1,1,2,2)$, respectively. We refer for details 
to \cite{HBB}, where  a possible breakdown mechanism
for the gauge group $SU(4)_{PS} \otimes [SU(2)]^4$ is discussed.
Needless to say, it is a quite complicated task to generate fermion
masses in general and we leave it for the future.

Experimentally, it is safe to assume
that any long-lived new quarks, if they exist, should 
have a mass  larger than
$200\,GeV$\cite{CDF,FHS}. For new leptons, the experimental 
lower bounds  are weaker ($45, 90\,GeV$ for  stable
and unstable neutral heavy 
leptons, respectively and $100\, GeV$ for the charged leptons
\cite{PDG}).

Now, the possible extra generations of ordinary fermions couple to 
the SM Higgs field. This normally means that they cannot be much heavier 
than, say, $200\gev$ and the SU(2) doublet partners have to be approximately 
degenerate in mass to be consistent with the electroweak precision studies.
We will assume that they have masses $\ord(250\gev)$. On the other hand, 
as the mirror fermions and the relevant Higgs system are singlets under
$SU(2)_L$, the latter restiction is absent. In fact as already found in  
\cite{HBB}, it is more favourable from the point of 
view of the RG analysis that the mirror fermion masses are close to 
$\tilde M$ so that their contributions  
to $K$ in (\ref{K}) can be neglected.

Finally, let us recall that in this model the ordinary quark and leptons are
coupled to each other by the heavy PS gauge bosons with masses $\ord(M)$ and 
electric charges $\pm 2/3$. The detailed presentation of the $SU(4)_{\rm PS}$
gauge boson sector can be found in \cite{HBB}, where also the implications of 
these quark-lepton couplings for very rare or forbidden decays have been 
analyzed. We will update this analysis in Sec. V.
\subsection{\boldmath{ $SU(4)_{\rm PS} \otimes [SU(2)]^3$}}
From Table 1, we see that $\sin^{2}\theta_{W}^{0}=1/3$ in this case
and one should have $C_{S}^{2} = 8/3$. What are the appropriate
fermion representations? As usual, the requirements are
simply that these representations are anomaly-free under
$SU(4)_{S} \otimes [SU(2)]^3$, and that they appear in a
sufficient number so as to ensure the equality of the three ``weak'' 
couplings above $\tilde M$. The most economical way to satisfy these 
requirements
is to have the following fermion content for each generation 
which also gives a rather
interesting physical interpretation of $[SU(2)]^3$:

(a) $(4,2,2,1)_L$,

(b) $(4,1,2,2)_R$,

(c) $(4,2,1,1)_L,\, (4,2,1,1)_R$,

(d) $(4,1,1,2)_L,\, (4,1,1,2)_R$.

This is clearly a situation in which one has mixed representations
of classes (i) and (ii). Before addressing the issues of charges,
let us first verify whether (a)-(d) are anomaly-free. If (a) and (b)
represent the same particles but with opposite chiralities, then
they are anomaly-free when combined. Also, (c) and (d)
are separately anomaly-free. In addition, the number of degrees of
freedom for (a)-(d) combined is exactly what one needs to
guarantee the equality of the $G_W$ couplings above $\tilde M$. 

The physical interpretation of $[SU(2)]^3$ is now clear, namely
\be
[G_W]_1= 
SU(2)_{L} \otimes SU(2)_{H} \otimes {SU(2)}_{R}~. 
\ee
As
we will show below $SU(2)_H$ is the ``horizontal'' gauge group
which links conventionally charged SM fermions to the
unconventionally charged ones. To clearly see these features,
let us write down explicitely the charge structure of the fermions
in (a)-(d). First we look at (a) and (b).

In accordance with (\ref{chargedis}), $Q_{W}$ for (a) and (b) 
is simply given by
\begin{equation}
Q_{W} = \left(
\begin{array}{cc}
0 & 1 \\ 
-1 & 0 
\end{array}
\right)\ \,
\label{QW22}
\end{equation}
with the columns and the rows representing $SU(2)_{L,R}$ and 
$SU(2)_{H}$ doublets, respectively.

With $C^2_S=8/3$, the electric charges of the quarks and leptons are 
then given by
\be\label{QQQL}
Q^i_q=1/3 + \tilde Q^i_W, \qquad Q^i_l=-1 + \tilde Q^i_W
\ee
and consequently
with (\ref{QW22}), these charges are
\begin{equation}
Q_q=\left(
\begin{array}{cc}
1/3 & 4/3 \\ 
-2/3 & 1/3 
\end{array}
\right)\ \,,
\label{Qqua}
\end{equation}
for the quarks and
\begin{equation}
Q_l=\left(
\begin{array}{cc}
-1 & 0 \\ 
-2 & -1 
\end{array}
\right)\ \,,
\label{Qlep}
\end{equation}
for the leptons.
Notice that one now has quarks and leptons with
unconventional charges, $4/3$ and $2$.

For (c) and (d), one has $Q_W= \pm 1/2$ as in (\ref{f11}). But since 
the charges of fermions are
still given by (\ref{QQQL}), one now has the following charge
assignments for the vector-like quarks and leptons:
$5/6, -1/6$ for the quarks, and $-1/2, -3/2$ for the leptons. 
These are the ``funny" charges mentioned in the previous section.
Let us remember that these are vector-like fermions and, therefore,
can possess large masses which are not connected to the electroweak scale,
nor to the scale of $SU(2)_R$ breaking. We shall come back to this
point in the RG analysis.

To facilitate the discussion, we now present the following
notations for the above quarks and leptons, for each generation.
We have (with the electric charges shown in parentheses):
\bes
\label{rep11}
\be
\psi^{q}_{L,R} = \left(
\begin{array}{c}
u(2/3)\\ 
d(-1/3)
\end{array}
\right)_{L,R}\ \,; 
\ee
\be
\tilde{Q}_{L,R} = \left(
\begin{array}{c}
\tilde{U}(4/3)\\ 
\tilde{D}(1/3)
\end{array}
\right)_{L,R}\ \,; 
\ee
\be
\psi^{l}_{L,R} = \left(
\begin{array}{c}
\nu(0) \\ 
l(-1)
\end{array}
\right)_{L,R}\ \,;
\ee
\be
\tilde{L}_{L,R} = \left(
\begin{array}{c}
\tilde{l}_{u}(-1)\\ 
\tilde{l}_{d}(-2)
\end{array}
\right)_{L,R}\ \,;
\ee
\be
\tilde{Q}^{\prime,\prime\prime}_{L,R} = \left(
\begin{array}{c}
\tilde{U}^{\prime,\prime\prime}(5/6)\\ 
\tilde{D}^{\prime,\prime\prime}(-1/6)
\end{array}
\right)_{L,R}\ \,;
\ee
\be
\tilde{L}^{\prime,\prime\prime}_{L,R} = \left(
\begin{array}{c}
\tilde{l}^{\prime,\prime\prime}_{u}(-1/2)\\ 
\tilde{l}^{\prime,\prime\prime}_{d}(-3/2)
\end{array}
\right)_{L,R}\ \,.
\ee
\ees

In order to put these $SU(2)$ doublets into representions 
(a)--(d), we note that
the following field transforms like a $\bar{2}$ which is
equivalent to a $2$ of $SU(2)_L$:
\begin{eqnarray}
i\tau_{2} \psi^{q,*}_{L,R} = \left(
\begin{array}{c}
d^{*}(1/3)\\ 
-u^{*}(-2/3)
\end{array}
\right)_{L,R}\ \, ,
\end{eqnarray}
with $\tau_{2}$ being an $SU(2)_{L,R}$ generator.

Using the above definitions, one can write
\be
\label{a}
(4,2,2,1)_L =[(i\tau_{2} \psi^{q,*}_{L},\, \tilde{Q}_{L}), 
(\tilde{L}_{L},\, \psi^{l}_{L})] \, ,
\ee
\be
\label{b}
(4,1,2,2)_R =[(i\tau_{2} \psi^{q,*}_{R},\, \tilde{Q}_{R}), 
(\tilde{L}_{R},\, \psi^{l}_{R})] \, ,
\ee
and
\be
\label{c}
(4,2,1,1)_{L,R} = [\tilde{Q}^{\prime}_{L,R},\,
\tilde{L}^{\prime}_{L,R}] \, ,
\ee
\be
\label{d}
(4,1,1,2)_{L,R} = [\tilde{Q}^{\prime\prime}_{L,R},\,
\tilde{L}^{\prime\prime}_{L,R}] \, .
\ee

Three remarks are in order here. 
\begin{itemize}
\item
First, the fermions in (\ref{c}, \ref{d}) are
vector-like and, in consequence, can have gauge-invariant bare masses
which can be much larger than the electroweak scale. 
\item
Second, the
placement of the quarks and leptons in (\ref{a}, \ref{b}) is such that there
are {\em no tree-level} transitions between ordinary quarks and leptons
mediated by the $SU(4)_{\rm PS}$ gauge bosons. 
Indeed, in contrast to the previous scenario the electric charges of the 
PS gauge bosons are now $\pm 4/3$ and as seen for instance in 
(\ref{a}), (\ref{Qqua}) and (\ref{Qlep}) these gauge bosons couple a 
left-handed ordinary anti-down-quark with charge $1/3$ to a new heavy $-1$ 
charge lepton and a left-handed ordinary charged lepton with charge $-1$ 
to a new heavy $1/3$ charge quark. Analogous comments apply to anti-up-quarks 
and neutrinos.
\item
Third, as seen explicitly in (\ref{Qqua}) and (\ref{Qlep}), the horizontal 
$SU(2)_H$ weak gauge bosons couple the ordinary quarks and leptons to 
new heavy quarks and leptons, respectively and consequently there are 
no dangerous tree level flavour changing neutral current (FCNC) 
transitions between the ordinary quarks and between the ordinary leptons 
mediated by the $SU(2)_H$ bosons.
\end{itemize}

As we shall see, the second property will
prevent rare decays such as $K_L \rightarrow \mu e$ from acquiring
large rates, even for the masses of the PS gauge bosons as low as 1 TeV.
Similar comments apply to horizontal $SU(2)_H$ gauge bosons with 
respect to FCNC transitions.

\subsection{\boldmath{ $SU(4)_{\rm PS} \otimes [SU(3)]^2$}}
In this scenario the weak gauge group is
\be
[G_W]_2= 
SU(3)_{L} \otimes SU(3)_{H} 
\ee
with the SM $SU(2)_L$ group being the subgroup of $SU(3)_L$. As 
we will show below the ``horizontal'' gauge group $SU(3)_H$ similarly 
to $SU(2)_H$ in the previous scenario links conventionally charged SM 
fermions to the unconventionally charged ones.

As we have discussed above, $\sin^{2}\theta_{W}^{0}=3/8$ in this
model and $C^2_S=8/3$ is required. The appropriate fermion representations 
that  are together 
anomaly free, are then
$(4,3,\bar{3})$ and $(4,\bar{3},3)$. The ``weak charge'' matrices
are now written as
\begin{equation}
Q_{W} = \left(
\begin{array}{ccc}
0 & 1 & 1\\ 
-1 & 0 & 0\\
-1  & 0 & 0
\end{array}
\right)\
\label{QW1}
\end{equation}
for $(4,3,\bar{3})$, and
\begin{equation}
Q_{W} = \left(
\begin{array}{ccc}
0 & -1 & -1\\ 
1 & 0 & 0\\
1  & 0 & 0
\end{array}
\right)\
\label{QW2}
\end{equation}
for $(4,\bar{3},3)$, both
with eigenvalues $0,\pm 1$. The charges for the fermions are given
by (\ref{QQQL}) as in the previous scenario, but as only representation of 
class ii) are present the fermions with ``funny" charges are absent. 
We will soon see that the rows in (\ref{QW1}) and (\ref{QW2}) correspond
to $SU(3)_L$ triplets with the $SU(2)_L$ doublets occupying the first two 
entries in these triplets. The columns in (\ref{QW1}) and (\ref{QW2}) 
correspond to $SU(3)_H$ triplets. 

From (\ref{fundamental2}, \ref{QW1}), the three fundamental
representations of $SU(3)_L$ 
 have the weak charge
distributions: $(0,1,1)$, $(-1,0,0)$, and $(-1,0,0)$. This
corresponds to the electric charge distributions:
$(1/3,4/3,4/3)$, $(-2/3,1/3,1/3)$, and $(-2/3,1/3,1/3)$
for the quarks and
$(-1,0,0)$, $(-2,-1,-1)$, and $(-2,-1,-1)$ for the leptons.
In short, each $(4,3,\bar{3})$ representation
will have the following fermion content:
\begin{eqnarray}
\Psi_1&=&([(1/3,4/3,4/3),(-1,0,0)], \nonumber \\ 
      & &[(-2/3,1/3,1/3),(-2,-1,-1)], \nonumber \\ 
      & &[(-2/3,1/3,1/3),(-2,-1,-1)])\,.
\label{psi1}
\end{eqnarray}
Similarly, each $(4,\bar{3},3)$
representation has the following fermion content:
\begin{eqnarray}
\Psi_2&=&([(1/3,-2/3,-2/3),(-1,-2,-2),] \nonumber \\ 
& & [(4/3,1/3,1/3),(0,-1,-1)], \nonumber \\
& & [(4/3,1/3,1/3),(0,-1,-1)])\,.
\label{psi2}
\end{eqnarray}

To appreciate the physical meaning of $\Psi_1$ and $\Psi_2$,
it is best to express them explicitely in terms of various
particles. In particular, we would like to clearly
distinguish fields which represent SM particles and
those which represent new kinds of particles.
For that purpose, we introduce left-handed Weyl
fields grouped together as $SU(2)_L$ doublets or singlets.
The electric charges are given in the parentheses. 
For the SM particles, we require, for each family, a left-handed
lepton doublet, a left-handed quark doublet, a right-handed
charged lepton, a right-handed up quark and a right-handed
down quark.

 Since it is convenient to put into a given representation
particles of the same chirality, we will make use, in subsequent
discussions, of the usual definition of a charge conjugate field:
\be
\label{conjugate}
\psi^{c}_{L,R} \equiv {\cal C} \psi^{q}_{L,R} {\cal C}^{-1} =
C \bar{\psi}_{R,L}^{T} \, ,
\ee
where $C = i \gamma^{2} \gamma^{0}$.

First, we start with the $(4,3,\bar{3})$ representation. 
We shall first list normal quarks and leptons, followed
by those which possess unusual electric charges. The notations
used below should not be confused with the ones used in
Section IIIC.
One has
\bes
\label{rep1}
\be
\psi^{q}_{L} = \left(
\begin{array}{c}
u(2/3)\\ 
d(-1/3)
\end{array}
\right)_L\ \,; \; d^{c}_L(1/3)=C \bar{d}^{T}_{R}\, .
\ee
\be
\psi^{l}_{L} = \left(
\begin{array}{c}
\nu(0) \\ 
l(-1)
\end{array}
\right)_L\ \,; \; \nu^{c}_L= C \bar{\nu}_{R}^{T} \, ;
\ee
\be
Q_{L} = \left(
\begin{array}{c}
U(-1/3)\\ 
D(-4/3)
\end{array}
\right)_L\ \,; \; D^{c}_{L}(4/3) = C \bar{D}^{T}_{R} \, ,
\ee
\be
L_{1L} = \left(
\begin{array}{c}
l_{u1}(2)\\ 
l_{d1}(1)
\end{array}
\right)_L\ \,; \; l^{c}_{d1,L}(-1)= C \bar{l}^{T}_{d1,R}\, ,
\ee
\be
L_{2L} = \left(
\begin{array}{c}
l_{u2}(2)\\ 
l_{d2}(1)
\end{array}
\right)_L\ \,,
\ee
\be
\tilde{\psi}^{q}_{L} = \left(
\begin{array}{c}
\tilde{u}(2/3)\\ 
\tilde{d}(-1/3)
\end{array}
\right)_L\ \,; \; d^{\prime}_L(-1/3), 
\ee
\be
l^{\prime}_L(+1) \, .
\ee
\ees

In the above, we have put particles in $SU(2)_L$ doublets and singlets.
To put these fields into the representation $(4,3,\bar{3})$, we
shall need the following $SU(2)_L$ doublets obtained from above:
\begin{eqnarray}
\label{doublet1}
&&i\tau_{2} L^{*}_{1L} = \left(
\begin{array}{c}
l_{d1}^{*}(-1)\\ 
-l_{u1}^{*}(-2)
\end{array}
\right)_{L}\ \, , \,
i\tau_{2} Q^{*}_{L} = \left(
\begin{array}{c}
D^{*}(4/3)\\ 
-U^{*}(1/3)
\end{array}
\right)_{L}\ \, , \nonumber \\
&&
i\tau_{2} \psi^{q,*}_{L} = \left(
\begin{array}{c}
d^{*}(1/3)\\ 
-u^{*}(-2/3)
\end{array}
\right)_{L}\ \, ,
i\tau_{2} \tilde{\psi}^{q,*}_{L} = \left(
\begin{array}{c}
\tilde{d}^{*}(1/3)\\ 
-\tilde{u}^{*}(-2/3)
\end{array}
\right)_{L}\ \, ,
\end{eqnarray}
where $\tau_2$ is a generator of $SU(2)_L$. One can now
write $(4,3,\bar{3})$ in terms of specific
fields, namely
\begin{eqnarray}
\Psi_1&=&([( i\tau_{2} Q^{*}_{L},D^{c}_{L}), 
           (\psi^{l}_{L}, \nu^{c}_L)], \nonumber \\ 
      & &[(i\tau_{2} \psi^{q,*}_{L},d^{c}_L), 
          (i\tau_{2} L^{*}_{1L},l^{c}_{d1,L})], \nonumber \\ 
      & &[(i\tau_{2} \tilde{\psi}^{q,*}_{L},d^{\prime,*}_L),
          (i\tau_{2} L^{*}_{2L},l^{\prime,*}_L)])\,.
\label{psi11}
\end{eqnarray}

From (\ref{psi11}), one can identify the SM fields, namely
$\psi^{l}_{L}, i\tau_{2} \psi^{q,*}_{L}, \nu^{c}_L,
 d^{c}_L$. However, this representation is incomplete in 
that the right-handed charged lepton and up-quark fields 
are missing. This is where the $(4,\bar{3},3)$ representation
comes in. The meaning of the non-SM fields appearing in
(\ref{psi11}) will be elucidated below.

For the $(4,\bar{3},3)$ representation, one can look at
(\ref{psi2}) to find the appropriate fields. To this end,
let us introduce
\bes
\label{rep2}
\be
\tilde{\psi}^{l}_{L,R} = \left(
\begin{array}{c}
\tilde{\nu}(0)\\ 
\tilde{l}(-1)
\end{array}
\right)_{L,R}\ \, ; \, l^{c}_{L}(+1) = C \bar{l}^{T}_{R} \, ;
\ee
\be
l^{\prime}_R (+1) \, ;
\ee
\be
\tilde{\psi}^{q}_{R} = \left(
\begin{array}{c}
\tilde{u}(+2/3)\\ 
\tilde{d}(-1/3)
\end{array}
\right)_R\ \,;
\ee
\be
L_{2R} = \left(
\begin{array}{c}
\l_{u2}(2)\\ 
l_{d2}(1)
\end{array}
\right)_{R}\ \, ; \,
l^{c}_{u1,L}(-2)= C \bar{l}^{T}_{u1,R} \, ;
\ee
\be
Q^{\prime}_{L,R} = \left(
\begin{array}{c}
U^{\prime}(-1/3)\\ 
D^{\prime}(-4/3)
\end{array}
\right)_{L,R}\ \,; \; U^{c}_{L}(1/3) = C 
\bar{U}^{T}_{R} \, ;
\ee
\be
d^{\prime}_R (-1/3) \, .
\ee
\ees

From the above equations, one can immediately identify the
following vector-like fields: $L_{2L,R}$, $Q^{\prime}_{L,R}$,
 $\tilde{\psi}^{l}_{L,R}$, $\tilde{\psi}^{q}_{L,R}$, 
$l^{\prime}_{L,R}$ and $d^{\prime}_{L,R}$.

Next, in order to match the charge assignments of (\ref{psi2}),
we define the following $SU(2)_L$ doublets, using
the ones defined in (\ref{doublet1}):
\bes
\label{doublet2}
\be
\tilde{\psi}^{l,c}_{L}\, = \, 
C \bar{\tilde{\psi}}^{l,T}_{R} \, = \, \left(
\begin{array}{c}
\tilde{\nu}^{c}_L (0)\\
\tilde{l}^{c}_L(+1)
\end{array}
\right)\ \, , 
\ee
\be
i\tau_2 L^{c}_{2L} \, = \, i\tau_2  \, 
C \bar{L}^{T}_{2R} \, = \, \left(
\begin{array}{c}
l^{c}_{d2,L}(-2) \\
-l^{c}_{u2,L}(-1)
\end{array}
\right)\ \, , 
\ee   
\be
i\tau_2 Q^{\prime,c}_{L} \, = \, i\tau_2  \, 
C \bar{Q^{\prime}}^{T}_{R} \, = \, \left(
\begin{array}{c}
D^{\prime,c}_L(4/3) \\
-U^{\prime,c}_L(1/3)
\end{array}
\right)\ \, ,
\ee 
\be
i\tau_2 Q^{\prime,*}_{L} \, = \, \left(
\begin{array}{c}
D^{\prime,*}_L(4/3) \\
-U^{\prime,*}_L(1/3)
\end{array}
\right)\ \, ,
\ee 
\be
i\tau_2 \tilde{\psi}^{q,c}_{L} \, = \, i\tau_2 \ \, 
C \bar{\tilde{\psi}}^{q,T}_{R} \, = \, \left(
\begin{array}{c}
\tilde{d}^{c}_L(1/3) \\
-\tilde{u}^{c}_L(-2/3)
\end{array}
\right)\ \, .
\ee 
\ees

The representation $(4,\bar{3},3)$ can now be written explicitely
as
\begin{eqnarray}
\Psi_2&=&([(i\tau_2 \tilde{\psi}^{q,c}_{L}, u^{c}_{L}),
         (i\tau_2 L^{c}_{2L},l^{c}_{u1,L}),] \nonumber \\ 
& & [(i\tau_2 Q^{\prime,c}_{L},U^{c}_{L}),
      (\tilde{\psi}^{l,c}_{L},l^{c}_{L})], \nonumber \\
   & & [(i\tau_{2} Q^{\prime,*}_{L},d^{\prime,c}_L),
         (\tilde{\psi}^{l,*}_{L},l^{\prime,c}_L )])\,.
\label{psi22}
\end{eqnarray}

Several remarks are in order here. First, the $(4,3,\bar{3})$ and
$(4,\bar{3},3)$ representations, as described by $\Psi_1$
and $\Psi_2$, together form an anomaly-free representation
of the group $SU(4)_{S} \otimes [SU(3)]^2$. Second, the
particle content described in (\ref{rep1}) and (\ref{rep2})
has the following features:
\begin{itemize}
\item
There are two types of families with SM transformations under
$SU(2)_L$, i.e. left-handed doublets and right-handed singlets:
one contains the SM quarks and leptons and the other one
contains unconventional quarks and leptons with charges up to
$4/3$ (for the quarks) and $2$ (for the leptons). 
 The unconventional fields are $Q_L$, $D^c_L$, $U^c_L$, $L_{1L}$, 
$l^c_{d1,L}$ and $l^c_{u1,L}$. The
(normal and unconventional) quarks and leptons couple to the
SM Higgs field. This normally means that their masses cannot
be much heavier than, say, $200\gev$.
\item
There are, in addition, two families of quarks and leptons, 
 $(\tilde\psi^q,\tilde\psi^l)_{L,R}$ and $(Q^\prime,L_2)_{L,R}$,
with normal and unconventional charges which are 
{\em vector-like} under $SU(2)_L$. This means that their
masses come from sources other than the SM Higgs field and
they can be much heavier than the first two types of families
mentioned above.
\item
Next, there are two {\em vector-like} $SU(2)_L$-singlets
with charge $+1$ for the lepton-like color-singlet ($l^\prime_{L,R}$) 
and charge
$-1/3$ for the quark-like color triplet ($d^\prime_{L,R}$). 
They also can acquire
large masses.
\end{itemize}
Finally as in the previous scenario we have two phenomenologically 
very relevant properties that can be clearly seen in 
(\ref{psi11}, \ref{psi22}):
\begin{itemize}
\item
The
placement of the quarks and leptons in (\ref{psi11}, \ref{psi22}) is such 
that there
are {\em no tree-level} transitions between ordinary quarks and leptons
mediated by the $SU(4)_{\rm PS}$ gauge bosons. 
Also here the electric charges of the 
PS gauge bosons are $\pm 4/3$. 
\item
The horizontal 
$SU(3)_H$ weak gauge bosons couple the ordinary quarks and leptons to 
new heavy quarks and leptons, respectively and consequently there are 
no dangerous tree level flavour changing neutral current (FCNC) 
transitions between the ordinary quarks and between the ordinary leptons 
mediated by the $SU(3)_H$ bosons.
\end{itemize}

\section{RG Analysis of \boldmath{$\sin^{2}\theta_{W}$}}
\subsection{Preliminaries}

In 1981 the values of $\sin^{2}\theta_{W}(M_Z^2)$ and $\alpha_s(M^2_Z)$ 
were rather poorly known. As of 2003 we know them with a very high 
precision as given in (\ref{newin1}) and (\ref{newin2}) with 
$\alpha_s(M^2_Z)$  substantially smaller than in 1981 so that 
the $\ord(\alpha/\alpha_s)$ correction in (\ref{sinsq2}) plays now a bigger 
role.
In this section we will update our 1981 renormalization group  analysis 
of ${\rm PUT}_0$ and generalize it to the additional scenarios considered 
in the previous section.

The master formula for $\sin^{2}\theta_{W}(M_Z^2)$ in (\ref{sinsq2}) 
has been obtained in the one-loop approximation, whereas the values 
of $\sin^{2}\theta_{W}(M_Z^2)_{\rm exp}$,  $\alpha_s(M^2_Z)$ and 
$\alpha(M^2_Z)$ have been extracted from various data including higher 
order QCD and electroweak corrections. Strictly speaking we should 
then generalize (\ref{sinsq2}) to include two-loop contributions.
This would be indispensible in the case of GUTS 
where $\mu$ varies from $M_Z$ to $10^{16}$ GeV and the change of the
gauge couplings in this range is substantial. On the other hand in the case 
of early unification, the changes of the couplings between $M_Z$ and 
$(\tilde M,M)$ that are in the TeV's range are rather small and the 
two-loop contributions to (\ref{sinsq2}) are insignificant.
In what follows we will therefore use the one-loop formula 
(\ref{sinsq2}), relegating the RG analysis at two-loop level to 
a future paper. 

While $M$ and $\tilde M$ differ in principle from each other, with 
$M\ge \tilde M$, we will first set $\tilde M=M$. Consequently
the last term in (\ref{sinsq2}) is absent and only the coefficient
$K$ has to be calculated. On the other hand in the scenarios considered, 
there are new particles with masses below $M$ and their contributions 
to (\ref{sinsq2}) have to be taken into account.
Now, as discussed in the previous section, all new particles with 
non-trivial properties under $SU(2)_L$ which are not vector-like 
cannot have masses much larger
than $200\gev$. In the RG analysis we will set all these masses to be equal
to a single scale $M_F$ with
\be\label{MF}
M_F=(250\pm 50)\gev
\ee
and we will assume that all the remaining new particles have masses 
very close to $M$ so that their contributions to (\ref{sinsq2}) can
be neglected.

Under these assumptions, the following replacement should be made in 
(\ref{sinsq2}):
\be\label{LOGS}
K\ln\frac{\tilde{M}}{M_Z}\to
K_{n_G=3}\ln\frac{M_F}{M_Z}+ K_{\rm total}\ln\frac{{M}}{M_F} 
\ee
where
\be
K_{n_G=3} = [b_1 - C_{W}^{2}b_2 - C_{S}^{2} b_3]_{n_G=3} \, ,
\end{equation}
with $b_i$'s receiving only contributions from the ordinary three 
generations ($n_G$) of quarks 
and leptons and the SM Higgs doublet. On the other hand
\be
K_{\rm total} = [b_1 - C_{W}^{2}b_2 - C_{S}^{2} b_3]_{\rm total} \, ,
\end{equation}
includes all particles with masses below $M$.

With $\tilde M=M$, $M_F$ given in (\ref{MF}),
  $\alpha_s(M^2_Z)$ and 
$\alpha(M^2_Z)$ known experimentally and $C^2_S$, $C^2_W$ and $b_i$ 
fixed (see below) in each scenario we can determine the value of $M$
that is consistent with the experimental value 
$\sin^{2}\theta_{W}(M_Z^2)_{\rm exp}$ in (\ref{newin2}). This is what 
we will do first. Subsequently we will analyze the general case with 
$\tilde M\le M$. In the next section we will investigate 
whether the values of $M$ determined here are consistent with bounds 
on rare decays.

\subsection{\boldmath{ $SU(4)_{\rm PS} \otimes [SU(2)]^4$}}
In this scenario
\be
 \sin^{2} \theta_W^0=\frac{1}{4}, \qquad C_W^2=3, \qquad C^2_S=\frac{2}{3}
\ee
and
\be\label{b1}
b_1=\frac{1}{48\pi^2}\left[\frac{20}{3}{n_G}+\frac{1}{2}\right],
\ee
\be\label{b2}
b_2=\frac{1}{48\pi^2}\left[4{n_G}+\frac{1}{2}-22 \right],
\ee
\be\label{b3}
b_3=\frac{1}{48\pi^2}\left[4{n_G}-33 \right]
\ee
with $n_G=3$ in $K_{n_G=3}$ and $n_G\ge 3$ in $K_{\rm total}$.
The ``1/2" is the contribution of the Higgs doublet.

We find then
\be
M\le 330\gev, \qquad  n_G=3
\ee
that is clearly excluded. Including new generations of ordinary fermions
 with masses 
$\ord(M_F)$ allows to increase $M$ as seen in the following formula 
\be\label{S1}
\sin^{2}\theta_{W}(M_Z^2)=0.2389-0.0065\ln\frac{M_F}{M_Z}
-0.0001 P\ln\frac{M}{M_F}~.
\ee
where
\be
P=87-8n_G~.
\ee
As the coefficient in front of the last logarithm in (\ref{S1}) must be 
very small in order to obtain the correct $\sin^{2}\theta_{W}(M_Z^2)$, 
the result for $M$ in this scenario is rather 
sensitive to the input parameters, in particular $n_G$ and $M_F$. However, 
requiring $M_F\ge 200\gev$ and $M\ge 800\gev$ we find the lowest acceptable 
value for $n_G$ to be $n_G=9$.

On the other hand making the model supersymmetric and setting as an example
the masses of all SUSY particles equal to $M_F$, one finds
\be\label{b1s}
[b_1]_{\rm total}=\frac{1}{48\pi^2}\left[10 {n_G}+3\right],
\ee
\be\label{b2s}
[b_2]_{\rm total}=\frac{1}{48\pi^2}\left[6{n_G}+3-18 \right],
\ee
\be\label{b3s}
[b_3]_{\rm total}=\frac{1}{48\pi^2}\left[6{n_G}-27 \right]~.
\ee
This gives the formula (\ref{S1})
with
\be
P=66-12 n_G
\ee
and the lowest acceptable value for $n_G$ to be $n_G=4$. For 
$n_G=3$ we find $M\le 550\gev$ that is excluded.

Whether this model is supersymmetric or not, the compatibility of this 
scenario with the experimental value of
$\sin^{2}\theta_{W}(M_Z^2)_{\rm exp}$ requires, for $M\ge 800\gev$, many
new particles around the $M_F$ scale.

The RG analysis of $SU(4)^2$ and $SU(8)$ proceeds in a similar manner 
but as these groups are very large we will not consider them further.

\subsection{\boldmath{ $SU(4)_{\rm PS} \otimes [SU(2)]^3$}}
In this scenario
\be
\sin^{2} \theta_W^0=\frac{1}{3}, \qquad C_W^2=2, \qquad C^2_S=\frac{8}{3}
\ee
and $[b_i]_{n_G=3}$ are simply given by (\ref{b1})--(\ref{b3}).
Above $M_F$ new generations of quarks and leptons with unconventional 
electric charges contribute and we find
\be\label{b1t}
[b_1]_{\rm total}=\frac{1}{48\pi^2}\left[\frac{20}{3}{n_G}+\frac{1}{2}
+\frac{116}{3}{n_G^{\rm new}}  \right],
\ee
\be\label{b2t}
[b_2]_{\rm total}=\frac{1}{48\pi^2}
\left[4({n_G}+n_G^{\rm new})+\frac{1}{2}-22 \right],
\ee
\be\label{b3t}
[b_3]_{\rm total}=\frac{1}{48\pi^2}
\left[4({n_G}+n_G^{\rm new})-33 \right]
\ee
with 
\be
n_G^{\rm new}=n_G~.
\ee
We note in particular the large contribution of the new fermions 
to $b_1$ that is related to high charges of these fermions.
This gives for $n_G=3$
\be\label{S2}
\sin^{2}\theta_{W}(M_Z^2)=0.2740-0.0132\ln\frac{M_F}{M_Z}
-0.0215\ln\frac{M}{M_F}~.
\ee
We observe that the coefficients of the logarithms are much larger 
than in the previous scenario and the correct value of 
$\sin^{2}\theta_{W}(M_Z^2)$ can be found with low unification scale and 
$n_G=3$ in spite of the much higher value of $\sin^2\theta^0_W$. 
Scanning $\alpha_s(M^2_Z)$ and $M_F$ in the ranges
(\ref{newin1}) and (\ref{MF}), respectively, and requiring
(at the two $\sigma$ level)
\be\label{range}
0.23083\le \sin^{2}\theta_{W}(M_Z^2) \le 0.23143
\ee
we find
\be\label{M1}
M=(1.00\pm 0.14)\tev, \qquad  n_G=3
\ee
with {\it lower} values for $n_G>3$. Thus in this scenario additional 
generations of ordinary quarks and leptons are disfavoured although 
$n_G=5$ would still give $M\ge 800\gev$. 
\subsection{\boldmath{ $SU(4)_{\rm PS} \otimes [SU(3)]^2$}}
In this scenario
\be
\sin^{2} \theta_W^0=\frac{3}{8}, 
\qquad C_W^2=\frac{5}{3}, \qquad C^2_S=\frac{8}{3}
\ee
and $b_i$ coefficients are the same as in the last scenario.
In this case (\ref{S2}) is replaced by
\be\label{S3}
\sin^{2}\theta_{W}(M_Z^2)=0.3083-0.0144\ln\frac{M_F}{M_Z}
-0.0243\ln\frac{M}{M_F}~.
\ee
and we find
\be\label{M2}
M=(3.30\pm 0.47)\tev, \qquad  n_G=3
\ee
with {\it lower} values for $n_G>3$. For instance for $n_G=4$ and $n_G=5$, 
M is found for the central values of input parameters in the ballpark of 
$3.0\tev$ and $2.6\tev$, respectively. 

\subsection{\boldmath{ $SU(4)_{\rm PS} \otimes [SU(2)]^2$}}
Finally, let us consider the original Pati-Salam model \cite{PASA}. Here
\be
\sin^{2} \theta_W^0=\frac{1}{2}, 
\qquad C_W^2=1, \qquad C^2_S=\frac{2}{3}
\ee
and $b_i$ coefficients are the same as in the 
$SU(4)_{\rm PS} \otimes [SU(2)]^4$ scenario. This
gives
\be
M\approx(5\cdot 10^{10})\tev, \qquad  n_G=3
\ee
with higher values for $n_G>3$. Clearly this model is not 
an early unification model.

\subsection{The case of \boldmath{$\tilde M\not=M$}}
Let us finally  consider the general case $\tilde M\le M$ with
$\tilde M\ge 800\gev$ as required by the lower limit of right-handed 
gauge boson masses in the case of $[SU(2)]^4$ and $[SU(2)]^3$ 
scenarios. The latter restriction is absent in the case of $SU(3)^2$ but
as we will see below in this case $\tilde M$ has to be above $1\tev$ if 
we want $M\le 10\tev$.

For $\tilde M\le M$  the last logarithm in
(\ref{LOGS}) is replaced as follows 
\be\label{LOGS2}
 K_{\rm total}\ln\frac{{M}}{M_F}\to
K_{\rm total}\ln\frac{{\tilde M}}{M_F}+ 
K'\ln\frac{{M}}{\tilde M}  
\ee
with $K^{'}$ defined in (\ref{K'}). 

Now, the values of $\tilde b$ and of 
$\tilde b_3$ relevant for the evolution of the couplings $\tilde g_S$ and 
$g_3$ for scales above $\tilde M$ include contributions from all fermions 
present in the model, that is also the vector-like ones. However, as 
$SU(3)_c$ and $U(1)_S$ are subgroups of $SU(4)_{\rm PS}$, the contributions 
of all fermions to $\tilde b$ and of 
$\tilde b_3$ are equal to each other at the one-loop level 
and consequently we find
\be
K^{'}=C^2_S \frac{33}{48\pi^2}
\ee
for all non-supersymmetric scenarios considered here
with $33$ replaced by $27$ in the case of Supersymmetry.

In the case of $PUT_0$ the factor $C^2_S 33=22$ in $K^{'}$ should be 
compared with $15$
present in $K_{\rm total}$ for $n_G=9$. Consequently the evolution between 
$\tilde M$ and $M$ is essentially the same as between $M_F$ and $\tilde M$ 
and making $\tilde M\not=M$ will not help to increase the value of $M$.
It will even lower it.

In the case of $PUT_1$ the factor $C^2_S 33=88$ in $K^{'}$ should be compared
with $311/2$ present in $K_{\rm total}$. Therefore lowering $\tilde M$ to 
$800\gev$ allows for central values of all parameters to increase $M$ from 
$1.0\tev$ in (\ref{M1}) to approximately $1.2\tev$.

In the case of $PUT_2$ the factor $C^2_S 33=88$ in $K^{'}$ should be compared
with $493/3$ present in $K_{\rm total}$. Therefore lowering $\tilde M$ to 
$800\gev$ allows for central values of all parameters to increase $M$ 
from $3.3\tev$ in (\ref{M2}) to as high as $9.9\tev$.

\begin{figure}[hbt]
\vspace{0.10in}
\centerline{
\epsfysize=3.1in
\epsffile{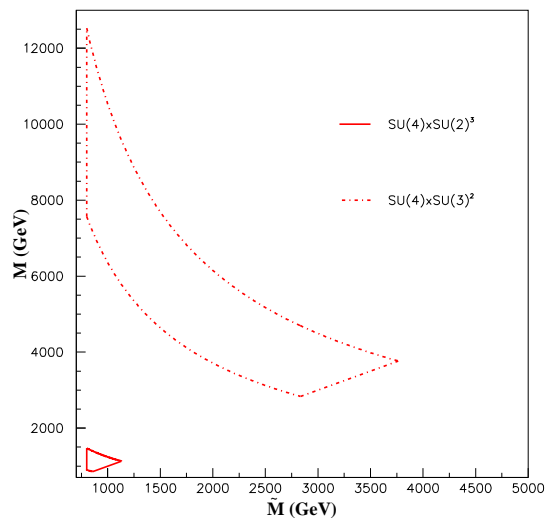}
}
\vspace{0.08in}
\caption{The allowed ranges for the $SU(2)^3$ and $SU(3)^2$ scenarios as 
discussed in the text.}
\label{fig:MM}
\end{figure}

In fig.~\ref{fig:MM} we show the allowed regions in the
space $(\tilde M,M)$ that have been obtained by varying $\alpha_S(M_Z^2)$, 
$M_F$ and $\sin^{2}\theta_{W}(M_Z^2)$ in the ranges (\ref{newin1}),
 (\ref{MF}) and (\ref{range}), respectively. For a given $\tilde M$, the 
maximal value of $M$ is found for the minimal $\sin^{2}\theta_{W}(M_Z^2)$
and maximal values of $M_F$ and $\alpha_S(M_Z^2)$. The
minimal value of $M$ is found for the maximal $\sin^{2}\theta_{W}(M_Z^2)$
and minimal values of $M_F$ and $\alpha_S(M_Z^2)$.
The vertical boundary lines at $\tilde M=800\gev$ have been set as discussed 
above and the boundery lines on the right represent the case $\tilde M=M$ 
considered previously. See the ranges in (\ref{M1}) and (\ref{M2}). 

We observe that even when $\tilde M\not=M$, the two scales have to be rather 
close to $1\tev$ in the $SU(2)^3$ scenario. On the other hand a much larger 
allowed region is obtained in the case of the $SU(3)^2$ scenario where 
$\tilde M$ and $M$ can differ even by an order of magnitude. However, we 
find that if $M$ is required to be less than $10\tev$, the scale $\tilde M$ 
has to be larger than  $\sim 1.1\,TeV$.

\subsection{Summary}
\begin{figure}[hbt]
\vspace{0.10in}
\centerline{
\epsfysize=3.1in
\epsffile{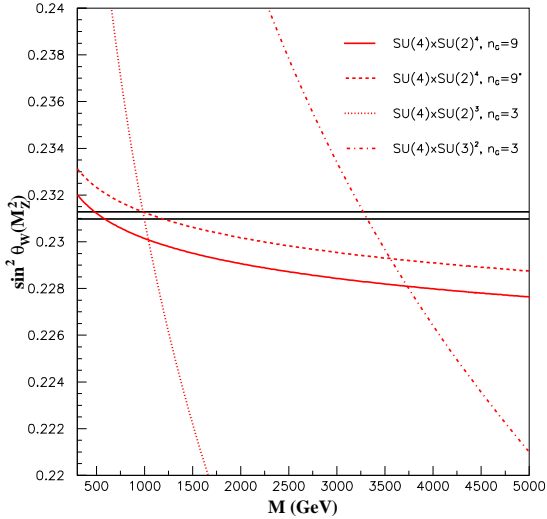}
}
\vspace{0.08in}
\caption{$\sin^{2}\theta_{W}(M_Z^2)$ as a function of $M$ in various
scenarios. The horizonal band represents the experimental value.
The dashed curve ($n_G = 9^{*}$) is obtained by using $M_F = 200 \gev$,
while the other three curves are obtained by using $M_F = 250\gev$.}
\label{fig:sin2}
\end{figure}

We observe that whereas the $SU(2)^4$ scenario requires new generations 
of ordinary quarks and leptons in order to be consistent with 
the experimental value of $\sin^{2}\theta_{W}(M_Z^2)$ and $M>800\gev$,  
in the case of the scenarios $SU(2)^3$ and $SU(3)^2$, the correct value of
$\sin^{2}\theta_{W}(M_Z^2)$ in the case of $\tilde M=M$ 
can be obtained with $n_G=3$ for $M\approx 1\tev$
and $M\approx 3.3\tev$, respectively. In fig.~\ref{fig:sin2} we show
$\sin^{2}\theta_{W}(M_Z^2)$ as a function of $M$ for the $SU(2)^4$ scenario 
with $n_G=9$ and for the scenarios $SU(2)^3$ and $SU(3)^2$ 
with $n_G=3$. To this end we have set 
$\alpha_s(M^2_Z)$ and $M_F$ to their central values.
The curve for the supersymmetric scenario $SU(2)^4$ with $n_G=4$ is rather 
similar to the non-supersymmetric case with $n_G=9$ shown in the figure.
The large sensitivity to $M_F$ in the case of the 
$SU(2)^4$ scenario is shown by the curve with $M_F=200\gev$.

Removing the equality $\tilde M=M$ and lowering $\tilde M$ to $800\gev$, 
has essentially no impact on the value of $M$ in the case of the $SU(2)^4$ 
scenario. An increase of $M$ by at most $300\gev$ is found in the case
of the $SU(2)^3$ scenario, implying that in this model $M$ and $\tilde M$ 
are forced to be of the same order of magnitude and in the ballpark of 
$1\tev$. On the other hand in the $SU(3)^2$ scenario $M$ can be by an order
of magnitude larger than $\tilde M$ and be as high as $12\tev$.
The allowed regions are shown in fig.~\ref{fig:MM}.

\section{On $K_{L} \rightarrow \mu \, e$}
\subsection{Preliminaries}
In our choice of $SU(4)_{\rm PS}$ as the strong group, we had already noticed in
\cite{HBB} that the heavy PS gauge bosons 
which connect quarks to
leptons can, in principle, induce the rare decay process
$K_{L} \rightarrow \mu \, e$. In the most naive version of the process,
$K_{L} \rightarrow \mu \, e$ can occur at tree-level
(only in the $SU(4)_{\rm PS} \otimes [SU(2)]^4$ case) 
if one assumes, as
we did in \cite{HBB}, some kind of ``kinship'' hypothesis such as
$d \leftrightarrow e$ and $s \leftrightarrow \mu$. That is no generation 
mixing. 
With this hypothesis,
we obtained an effective Lagrangian for the subprocess
$d\,+\,\mu \rightarrow e\,+\,s$ of the form
\be
\label{Leff}
{\cal L}^{d\mu \rightarrow es}_{eff} = \sqrt{2}G_{S} \sum_{i=1}^{3}
(\bar{d}_{i}\gamma_{\mu}e\bar{\mu}\gamma^{\mu}s_{i}\, + \,h.c.) \, ,
\ee
where the sum is over color and where
\be
\label{G_S}
g_{S}^{2}/2\,m_{G}^{2} = \sqrt{2}G_{S}.
\ee
In (\ref{G_S}), the quantity $m_G$ represents a typical mass of
the PS gauge bosons and is 
comparable to the scale $M$. 

In \cite{HBB} we have made the estimate of the branching ratio for 
$K_L\to \mu^\pm e^\mp$ by comparing this decay with $K_L\to\mu\bar\mu$. 
However, it will be more convenient to calculate $Br(K_L\to \mu^\pm e^\mp)$ 
directly. Making the Fierz transformation in (\ref{Leff}) and neglecting 
the axial-vector-current contribution as in \cite{HBB}, we find the 
amplitude
\be
A(K_L\to \mu^\pm e^\mp)=iF_K G_S \frac{m_K^2}{m_s+m_d}
[(\bar\mu\gamma_5e)+(\bar e\gamma_5\mu)]
\ee
where $m_K$ is the kaon mass, $F_K$ the kaon decay constant and $m_{s,d}$ are
the current quark masses. Neglecting the electron mass we find
\begin{eqnarray}
\label{BR}
Br(K_L\to \mu^\pm e^\mp)&=& \frac{\pi}{2}\frac{\alpha^2}{m_G^4} m_K F^2_K 
\tau(K_L)\sqrt{1-\frac{m_\mu^2}{m_K^2}} \nonumber \\
& & \times \left[\frac{m_K^2}{m_s+m_d}\right]^2
\end{eqnarray}
Using $F_K=160\mev$, $m_s+m_d=140\mev$ and the values for $m_K$, $\tau(K_L)$ 
and $m_\mu$ from \cite{PDG} we find
\begin{eqnarray}
Br(K_L\to \mu^\pm e^\mp)&=& 4.7\cdot 10^{-12}
\left(\frac{\alpha_S(m_G)}{0.1}\right)^2 \nonumber \\
& &  \times \left[\frac{1.8\cdot 10^3\tev}{m_G}\right]^4 
\end{eqnarray}
to be compared with the experimental bound \cite{PDG}
\be
\label{Kmue}
Br(K_L \rightarrow \mu e) < 4.7 \times 10^{-12}.
\ee

Now, 
$\alpha_{S}(m_G) = \alpha_{3}(m_G)$ 
and as the presence of new particles at scales lower than $m_G$ slows 
down the running of the QCD coupling constant, $\alpha_{3}(m_G)$
with $m_G=\ord( 1\tev)$ is not siginificantly different from 0.1.
We conclude then that in a scenario with no generation mixing and tree 
level contributions, the branching ratio
$Br(K_L\to \mu^\pm e^\mp)$ with $m_G=\ord( 1\tev)$ violates the experimental 
bound by at least {\em thirteen} orders of magnitude!

Let us then consider the presence of possible mixing
among generations. To be correct, we first denote the $T_{3L}=-1/2$
quarks by $D_{0}= (d_0, s_0, b_0)$ and similarly by 
$L_{0}=(e_0, \mu_0, \tau_0)$ for the leptons, with the subscripts $0$
referring to the eigenstates before mass mixing. A typical
$SU(4)/(SU(3) \otimes U(1)_{B-L})$ current would be of the form
$J^{\mu}_{LQ} = \bar{D}_{0} \gamma^{\mu} L_{0}$. 
Notice that this discussion only applies to the case
$SU(4)_{\rm PS} \otimes [SU(2)]^4$ where tree-level
SM lepto-quark transitions can occur.
If we now
diagonalize the mass matrices for the down quark and
for the charged lepton sectors, we can express $D_0$ and
$L_0$ in terms of the mass eigenstates as follows:
$D= U_{D} D_0$ and $L = U_{L} L_0$. The above current
can be rewritten as 
$J^{\mu}_{LQ} = \bar{D} \gamma^{\mu}U_{D} U^{-1}_{L} L$.
One now has the quark-lepton mixing matrix
$V_{LQ} = U_{D} U^{-1}_{L}$ involved in all quark lepton
transitions. In consequence, what should appear on the right-hand
sides of (\ref{Leff}) and (\ref{BR}) are 
extra factors  
$V^*_{ed}V_{\mu s}$ and $|V^*_{ed}V_{\mu s}|^2$, respectively. 
Here $V_{ed}$ and $V_{\mu s}$ are matrix
elements of $V_{LQ}$.

In the absence of a convincing model
of fermion masses, there is no reason to rule out 
the possibility
that the mixing coefficient $|V_{ed}V_{\mu s}|^2$ could be
of order $10^{-13}$, but such a very strong suppression 
appears rather strange and unnatural.
 Moreover, as $V_{LQ}$ is a unitary matrix  
not all of its elements can be set to zero and consequently even 
if the $K_L\to\mu e$ bound can be satisfied in this manner, other 
elements of $V_{LQ}$ that are relevant for lepton flavour violation
in B decays could be too large. Clearly the presence of more than 
three generations and consequently of many free parameters in $V_{LQ}$ 
could help but such a fine tunning in essentially all processes is 
rather ad hoc.

We conclude therefore that an early unification of quark and leptons 
requires either the absence of tree level contributions to 
$K_L\to\mu e$ and to analogous very rare decays or the presence of new
suppression mechanism in addition to $|V^*_{ed}V_{\mu s}|^2$  
considered above.

We shall now discuss the implication of these findings on the 
three candidates presented
in the previous section, namely $SU(4)_{\rm PS} \otimes [SU(2)]^4$,
$SU(4)_{\rm PS} \otimes [SU(2)]^3$, and $SU(4)_{\rm PS} \otimes [SU(3)]^2$.

\subsection{\boldmath{ $SU(4)_{\rm PS} \otimes [SU(2)]^4$}}
In this scenario, 
the decay $K_L\to \mu e$ takes place at tree level and
the RG analysis above
has shown that the PUT scale is typically around $1\tev$ or less
in order to agree with the experimental value for
$\sin^{2}\theta_{W}(M_Z^2)$. Consequently, as just discussed, 
this scenario is ruled out unless additional suppression mechanisms 
in addition to $|V^*_{ed}V_{\mu s}|^2$ can be invoked.

This could come from aspects of physics of Large Extra
Dimensions for example. One could add, for instance, an extra
spatial dimension (for the purpose at hand) and denote it,
for simplicity, by $y$. It has been shown that the compactification
of this extra dimension on an orbifold $S_{1}/Z_{2}$ gives
rise to chiral zero modes in four dimensions \cite{georgi}. 
 In \cite{CHHAPE}, it was proposed that $SU(4)_{\rm PS}$
is broken by boundary conditions. As a consequence, a quartet
which contains a quark and a lepton can only have one chiral
zero mode which could be either a quark or a lepton, with the
other one being a heavy partner. Since SM particles are
supposed to be chiral zero modes in four dimensions, they cannot
belong to the same quartet. Therefore there is no transition
between SM quarks and leptons via the PS  
gauge bosons at tree level, and $M$ can be as low as a few TeV's.
Another possibility is the following scenario.
The interaction of these chiral zero modes with a
background scalar field which has a kink solution along
the extra dimension has the effect of localizing these
chiral zero modes at various locations along $y$. These
chiral zero modes would represent the quarks and leptons
of the SM. An effective interaction in four dimensions
which involves a quark and a lepton, such as the leptoquark
transition generated by the PS 
gauge bosons, will contain a factor
\be
\label{overlap}
C_{ql} = \int \xi_{q}(y) \xi_{l}(y) dy \, ,
\ee
in the effective coupling, where $\xi_{q}(y)$ and
$\xi_{l}(y)$ represent the wave functions along $y$
of the quark and lepton chiral zero modes respectively.
When the quarks and leptons are localized far away from
each other along $y$, the factor $C_{ql}$ can be exponentially
small \cite{arkani}. If this scenario is correct then the bound
(\ref{Kmue}) can easily be satisfied for this model if
$|V_{ed}V_{\mu s} C_{de} C_{s\mu}|^2$ is of order $10^{-13}$.
Even if $|V_{ed}V_{\mu s}|^2$ were of the order of
unity, it is not hard to arrange for $|C_{de} C_{s\mu}|^2$ to be of
order $10^{-13}$, i.e. for $|C_{de} C_{s\mu}| \sim 10^{-6}$.

We observe then that the constraint
from $K_{L} \rightarrow \mu \, e$ has severe implications
on the $SU(4)_{S} \otimes [SU(2)]^4$  model because
of the low  PUT scale as required by
 the fit to the value of $\sin^{2}\theta_{W}(M_Z^2)$. It
implies either or both of the following scenarios: 1) The mass
matrices are such that $|V_{ed}V_{\mu s}|^2$ is very small;
and/or 2) The existence of a supression mechanism coming from
the physics of Large Extra Dimensions.

\subsection{\boldmath{ $SU(4)_{\rm PS} \otimes [SU(2)]^3$}}
As we have seen in Section \ref{content}, the particle content
of this group is rather interesting. The SM fermions belong
to $(4,2,2,1)_L =[(i\tau_{2} \psi^{q,*}_{L},\, \tilde{Q}^{\prime}_{L}), 
(\tilde{L}_{L},\, \psi^{l}_{L})]$ and
$(4,1,2,2)_R =[(i\tau_{2} \psi^{q,*}_{R},\, \tilde{Q}^{\prime}_{R}), 
(\tilde{L}_{R},\, \psi^{l}_{R})]$. From this fermion content,
one can see that the $SU(4)/(SU(3) \otimes U(1)_{B-L})$ gauge bosons 
with electric charges $\pm 4/3$ link the normal
quarks $i\tau_{2} \psi^{q,*}_{L,R}$ with the higher charged leptons
$\tilde{L}_{L,R}$, and the normal leptons $\psi^{l}_{L,R}$ with
the higher charged quarks $\tilde{Q}^{\prime}_{L,R}$. What this
implies is that, at tree level, there is NO transition between
normal quarks and normal leptons. 
However, it can occur at the one-loop level through a box diagram with 
two PS boson exchanges ($M_{\rm PS}=\ord(M))$ and new heavy quarks 
($\tilde Q$) and new heavy leptons ($\tilde L$) that have masses $\ord(M_F)$ 
with $M_F$ given in (\ref{MF}). $\tilde Q$ and $\tilde L$ appear in three 
generations and the mixing between these generations is given by $3\times3$ 
matrices to be denoted by $U$ and $V$, repectively. In the case of 
degenerate masses of $\tilde Q_i$ and $\tilde L_i$ the GIM mechanism is at 
work and the decay $K_L\to \mu e$ is absent. However, GIM mechanism 
remains to be powerful also when the masses are non-degenerate but all in 
the range $200-300\gev$. In this case it provides a suppression factor 
of $\ord(10^{-4})$ at the level of the branching ratio. With the typical 
loop factor $(16\pi^2)^{-2}\approx 4\cdot 10^{-5}$, the upper bound on 
the relevant mixing factors $|V_{id}V_{is}^*|^2|U_{jd}U_{js}^*|^2$   
coming from $K_L\to \mu e$ amounts then roughly to $\ord(10^{-4})$ and 
can be easily satisfied.

A detailed presentation of this calculation and the analysis of FCNC 
processes mediated by the $SU(2)_H$ bosons 
is beyond the scope of this paper and will be presented elsewhere but 
this discussion shows that in this scenario, the low unification scale 
required by the value of $\sin^{2}\theta_{W}(M_Z^2)$ is consistent 
with the present upper bound on $K_L\to\mu e$ and does note pose any 
problems with FCNC transitions at present.  

\subsection{\boldmath{ $SU(4)_{\rm PS} \otimes [SU(3)]^2$}}
The constraint coming from $K_{L} \rightarrow \mu \, e$ in this
model is very similar to the previous one. A look at the fermion
content, as shown in (\ref{psi11},\ref{psi22}), reveals that
the PS gauge bosons 
once more link normal quarks and
leptons to their higher charged counterparts. As a result,
there is no tree level contribution to $K_{L} \rightarrow \mu \, e$.
Again this process will occur at one loop, with an analysis similar
to the one mentioned above.
\section{Comparison with the Literature}
In order to make an assessment of our work and compare it with
recent attempts at ``low scale'' unification, we summarize below
the essential results which were presented above.
The three ``simplest'' cadidates for Petite Unification- a possible
nickname could be ``Tevunification''- are
$SU(4)_{\rm PS} \otimes [SU(2)]^4$, $SU(4)_{\rm PS} \otimes [SU(2)]^3$,
and $SU(4)_{\rm PS} \otimes [SU(3)]^2$. As mentioned at various
places in the paper, the philosophy of our Petite Unification
is to have a unification scale $M \leq 1000\, TeV$ and
preferably $M \leq 10\tev$.

\begin{itemize}
\item
${\rm PUT}_0=SU(4)_{\rm PS} \otimes SU(2)_{L} \otimes
SU(2)_{R} \otimes \tilde{SU(2)}_{L} \otimes \tilde{SU(2)}_{R}:$

This is the favorite scenario in our 1981 paper \cite{HBB}.
This model has only quarks and leptons (including possible new ones) 
having standard electric charges.
In our update of various numerical results, the conclusions drawn
from our analysis can be summarized as follows. In order to
obtain the correct value of $\sin^{2}\theta_{W}(M_{Z}^{2})$ and
requiring that $M \sim 1\tev$, our RG analysis
(assuming $\tilde{M}=M$) reveals that
we need  at least nine generations ($n_G = 9$), with the new generations 
having masses of 
order $250\gev$, or $n_G = 4$ if we include supersymmetry. 
In our RG analysis, the main important assumption which is made 
is that the masses of all new particles
are taken to be of order $250\gev$. No additional assumptions
are made about extra new physics other than Petite Unification
above the scale $M$ at this stage.

However, this scenario with a PUT scale of order $1 \tev$
suffers from the problem with the branching ratio for the 
process $K_{L} \rightarrow \mu \, e$ which in this scenario
can occur at tree level. Several possible remedies were
discussed above, in particular in the context of the physics of 
Large Extra Dimensions.
\item
${\rm PUT}_1=SU(4)_{\rm PS} \otimes SU(2)_{L} \otimes
SU(2)_{H} \otimes SU(2)_{R}:$

In this model the PUT scale is required to be $M \sim 1\tev$. 
In addition to the standard three generations of quark and leptons, new 
three generations of unconventional quarks and leptons
with charges up to $4/3$ (for quarks) and 2 (for leptons) 
and masses $\ord(250\gev)$ are automatically present.  
The horizontal groups $SU(2)_H$ connects the standard
fermions with the unconventional ones.
In addition,
there are also very heavy vector-like particles which, however,
are irrelevant to the phenomenology discussed in this paper.
Furthermore, in
this model, the process $K_{L} \rightarrow \mu \, e$ is forbidden
at tree level and appears only at the one-loop level. In consequence,
despite the appearance of a low PUT scale, the constraint from
$K_{L} \rightarrow \mu \, e$ can easily be satisfied, in contrast with
the $SU(2)^4$ scenario. No additional new physics such as 
Large Extra Dimensions is needed at this stage.
\item
${\rm PUT}_2=SU(4)_{\rm PS} \otimes SU(3)_L \otimes SU(3)_H:$

In this model the PUT scale is required to be in the range
$M \sim 3.3-10\tev$. Here,
the horizontal groups $SU(3)_H$ connects the standard
fermions with the unconventional ones.
It also contains new higher charged quarks and leptons
with masses as in the $SU(2)^3$ scenario. Also,
the process $K_{L} \rightarrow \mu \, e$ occurs only at one loop,
and the experimental bound for this decay can be easily satisfied as well. 
Again, no additional
new physics is needed at this stage.
\end{itemize}

In summary, $PUT_1$ and $PUT_2$ are able to predict
$\sin^{2}\theta_{W}(M_{Z}^{2})$ and to satisfy the
constraint on $K_{L} \rightarrow \mu \, e$ {\em within
the perturbative regime}. The offshoot of this is
the prediction of the existence of 
three generations of unconventional quarks and leptons
with charges up to $4/3$ (for quarks) and 2 (for leptons) 
and masses $\ord(250\gev)$.

Having briefly summarized the results of our three ``favorite'' scenarios,
we are now ready to make a comparison with the literature (surely an
incomplete task). In particular, we would like to compare our results with
those of \cite{CHHAPE} and \cite{DIKA}, whose main focus was
to derive $\sin^2 \theta_{W}$.

Ref. \cite{CHHAPE} basically generalized our 
$SU(4)_{\rm PS} \otimes [SU(2)]^4$ model of 1981 to Large Extra Dimensions.
This paper was motivated by the possibility of a TeV scale
unification. The first goal there was to obtain a reasonable
estimate for $\sin^{2}\theta_{W}(M_{Z}^{2})$ for a unification
scale of $\ord(1\tev)$. 
The second goal was to prevent
the process $K_{L} \rightarrow \mu \, e$ from acquiring
a large branching ratio due to the low unification scale.
To reach the first goal, a number of assumptions were made:
the size of the cut-off scale where the regime of strong
couplings set in (one might wonder whether
or not the leading log approximation is still valid), 
the size of the tree-level boundary corrections,
the contribution from the relative running of the $SU(2)$
gauge couplings above the compactification scale. In particular, this
last assumption, which is very model-dependent, is crucial
in obtaining an agreement with data. 
 We have checked that when supersymmetric contributions to the running 
of coupling constants are switched on only above $200\gev$ and not at $M_Z$ 
as done in \cite{CHHAPE} it is not possible to obtain acceptable solutions 
for the situation in which the $SU(2)$ gauge couplings
run parallel to each other as the correct value of the weak mixing angle
would require with $n_G=3$ 
a compactification 
scale significantly lower than $1\tev$. On the other hand in a model 
in which the breakdown of gauge symmetries is accomplished 
by using boundary conditions, the authors of \cite{CHHAPE} find a positive
contribution to   $\sin^{2}\theta_{W}(M_{Z}^{2})$ from scales higher than the 
compactification scale and the correct value of the mixing angle can be 
found for the compactification scale $\ord(2\tev)$.
In summary, the actual ``prediction''
for $\sin^{2}\theta_{W}(M_{Z}^{2})$ in this model depends crucially
on the assumptions made about various details of the physics
of Large Extra Dimensions. The second goal mentioned above
is achieved by the orbifold boundary conditions which split
a quartet of $SU(4)_{PS}$ into zero and non-zero modes. Since
the SM particles are supposed to be surviving zero modes in four
dimensions, ordinary quarks and leptons cannot be in the same quartet, 
similarly to the case of the $SU(2)^3$ and $SU(3)^2$ models 
considered here.
Consequently there are no tree-level transitions between SM quarks and
leptons and the $SU(4)/(SU(3) \otimes U(1)_{B-L})$ 
gauge bosons can be relatively ``light'' ($\ord(1\tev)$) without violating 
the upper bound on the rate of $K_L\to\mu e$. This
model predicts heavy copies of the SM particles with masses
of $\ord(1\tev)$.

Ref.\cite{DIKA} proposed to extend the Standard Model
$SU(2)_L \otimes U(1)_Y$ to $SU(3) \otimes SU(2) \otimes
U(1)$ at some scale $M$ of $\ord(1\tev)$. In this model,
$SU(3) \otimes SU(2) \otimes U(1) \rightarrow 
SU(2)_L \otimes U(1)_Y$ at $M$ which gives the following
relations between the couplings of the SM and its parent
group, namely 
\be
\frac{1}{g_2^2} = \frac{1}{g_{3}^2} + \frac{1}{\tilde{g}^2},
\quad
\frac{1}{g^{\prime,2}} = \frac{3}{g_{3}^2} + \frac{1}{\tilde{g}^{\prime,2}}
\ee
where the couplings on the right-hand side of
these equations belong to those of the parent group
while those on the left-hand side are those of the SM.
In the limit $\tilde{g},\tilde{g}^{\prime} \rightarrow
\infty$ ( the exact $SU(3)$ limit), one can easily 
derive $\sin^{2} \theta_W^0 =1/4$. Using the RG equations 
for $g_2$ and $g^{\prime}$ to match the value of
$\sin^{2} \theta_W$ at $M_Z$, Dimopoulos and Kaplan
obtained a value for the unification scale
$M_0 = 3.75\, TeV$ in the limit $\tilde{g},\tilde{g}^{\prime} 
\rightarrow \infty$. As mentioned in \cite{CHHAPE}, this
prediction is not precise because of these assumptions.
Once more, one is facing the problem with strong couplings.
Furthermore, unlike the case with the Pati-Salam group
or with the quintessential Grand Unified Theories, there is
no charge quantization in this scenario. However it is similar
in spirit to our 1981 paper \cite{HBB} in that 
$\sin^{2} \theta_W^0$ is determined entirely from the weak group
although two of the groups in \cite{DIKA} are not so weak 
after all. Notice that the exact $SU(3)$ limit of \cite{DIKA}
giving $\sin^{2} \theta_W^0 =1/4$ is similar to our case
of $G_W = SU(3)$ (with two doubly charged gauge bosons) 
as discussed in \cite{HBB} and mentioned in Section IIB.
In our case, this is ruled out by $\sin^2\theta_W(M_Z^2)$.

 Finally, in addition to \cite{HBB}, there are another two papers
within the past three years which dealt with $SU(3) \otimes SU(3)^2$
\cite{kim} and $SU(4)\otimes SU(2)^3$ \cite{triant} in a very different
context.

\section{Conclusions}
We have revived our previous paper \cite{HBB} that provided a general
discussion of an early quark-lepton unification characterized by the 
gauge group 
$G_S\otimes G_W$. As a byproduct we have presented a simple formula 
(\ref{simple2}) for $\sin^2\theta_W^0$ in the case of $G_W=SU(N)^k$
that is equivalent to the formula in \cite{HBB} but is more transparent.

During the last twenty two years the experimental value for
$\sin^2\theta_W(M_Z^2)$ became very precise and the value of 
$\alpha_s(M^2_Z)$ became not only more precise but also significantly 
smaller. As a result of these changes, our favourite 1981 scenario, 
$SU(4)_{\rm PS} \otimes [SU(2)]^4$, cannot be made consistent simultaneously
 with 
the data for $\alpha_s(M^2_Z)$ and the lower bound on the masses of 
right-handed gauge bosons unless six new generations of ordinary 
quarks and leptons are present. However, with the very low unification 
scale $\ord(1\tev)$, the improved experimental upper bound on $K_L\to\mu e$ 
is violated in this model by many orders of magnitude unless new, not
always natural, strong suppression factors are invoked.

Fortunately, we have found two new petite unification models for which the 
situation is much more favourable. These are the models based on the 
groups $SU(4)_{\rm PS} \otimes [SU(2)]^3$ and 
$SU(4)_{\rm PS} \otimes [SU(3)]^2$, of which the first one is more appealing 
in view of its simpler fermion content. The interesting properties 
of these models, described already briefly in Sec.I and in detail in 
Sec. III-IV are as follows:
\begin{itemize}
\item
The correct value of $\sin^2\theta_W(M_Z^2)$ with the unification scale in
the ballpark of $1\tev$ and $3-10\tev$, respectively.
\item
The absence of tree level lepton flavour violation and of tree level FCNC 
processes. These transitions are generated at one-loop through the 
exchanges of the heavy PS gauge bosons, new heavy quarks and leptons with 
unconventional electric charges (up to $4/3$ for quarks and $2$ for leptons )
and through the exchanges of ``horizontal" weak gauge bosons that couple 
the ordinary quarks and leptons with these new heavy fermions. Due to the 
GIM--like mechanism the bound on $K_L\to \mu e$ can easily be satisfied and 
the FCNC processes put under control.
\end{itemize}

The rich phenomenology resulting in these two new scenarios will be
presented in detail in a forthcoming paper.

Finally, we would like to stress the fact that the physics of our
two scenarios, $PUT_1$ and $PUT_2$, stands on its own regardless of
whether or not TeV-scale Large Extra Dimensions exist. Even if
they do exist, the predictions of $PUT_1$ and $PUT_2$ would
be {\em independent} of the details of the physics of Large Extra 
Dimensions.

\begin{acknowledgments}
AJB was partially supported by the German `Bundesministerium f\"ur 
Bildung und Forschung' under contract 05HT1WOA3 and by the 
`Deutsche Forschungsgemeinschaft' (DFG) under contract Bu.706/1-1.
 PQH  is supported in parts by the US Department
of Energy under grant No. DE-A505-89ER40518. 
\end{acknowledgments}


%
%


%
%

\end{document}